\documentclass[journal,a4paper]{IEEEtran}

\usepackage{amsmath,amsfonts}
\usepackage{algorithmic}
\usepackage{algorithm}
\usepackage{array}
\usepackage{textcomp}
\usepackage{stfloats}
\usepackage{url}
\usepackage{verbatim}
\usepackage{graphicx}
\usepackage{cite}
\usepackage{amsmath,amssymb,amsfonts}
\usepackage{algorithmic}
\usepackage{graphicx}
\usepackage{svg}
\usepackage{hyperref}
\usepackage[export]{adjustbox}
\usepackage{multirow}
\usepackage{subcaption}
\usepackage{fancyhdr}
\usepackage{textcomp}
\usepackage{xcolor}
\usepackage{siunitx}
\usepackage{booktabs}
\usepackage[textsize=tiny, textwidth=1.35cm]{todonotes}
\usepackage{selectp}

\begin{document}

\title{Position Tracking using Likelihood Modeling of Channel Features with Gaussian Processes}


\author{
    Sebastian Kram\IEEEauthorrefmark{1},
    Christopher Kraus\IEEEauthorrefmark{1},
    Tobias Feigl\IEEEauthorrefmark{1},
    Maximilian Stahlke\IEEEauthorrefmark{1},
    J\"org Robert\IEEEauthorrefmark{2}, and\\
    Christopher Mutschler\IEEEauthorrefmark{1}\\
    {\tt\footnotesize\{ sebastian.kram, krauscr, tobias.feigl, maximilian.stahlke, christopher.mutschler \}@iis.fraunhofer.de}\\
    {\tt\footnotesize joerg.robert@tu-ilmenau.de}\\
    {\vspace{0.125cm}}
    \and
    \IEEEauthorblockA{\IEEEauthorrefmark{1} Fraunhofer IIS, Fraunhofer Institute for Integrated Circuits IIS,
        Nuremberg, Germany\\}
    \and
    \IEEEauthorblockA{\IEEEauthorrefmark{2} Technical University of Ilmenau, Ilmenau, Germany}
}

\maketitle

\thispagestyle{fancy}

\begin{abstract}
Recent localization frameworks exploit spatial information of complex channel measurements (CMs) to estimate accurate positions even in multipath propagation scenarios. State-of-the art CM fingerprinting(FP)-based methods employ convolutional neural networks (CNN) to extract the spatial information. However, they need spatially dense data sets (associated with high acquisition and maintenance efforts) to work well -- which is rarely the case in practical applications. If such data is not available (or its quality is low), we cannot compensate the performance degradation of CNN-based FP as they do not provide statistical position estimates, which prevents a fusion with other sources of information on the observation level.

We propose a novel localization framework that adapts well to sparse datasets that only contain CMs of specific areas within the environment with strong multipath propagation. Our framework compresses CMs into informative features to unravel spatial information. It then regresses Gaussian processes (GPs) for each of them, which imply statistical observation models based on distance-dependent covariance kernels. 
Our framework combines the trained GPs with line-of-sight ranges and a dynamics model in a particle filter. Our measurements show that our approach outperforms state-of-the-art CNN fingerprinting (0.52 m vs. 1.3 m MAE) on spatially sparse data collected in a realistic industrial indoor environment.
\end{abstract}


\section{Introduction}
\label{sec:introduction}

Classical radio-frequency (RF) positioning mostly relies on the extraction of signal propagation times (or their differences) and a subsequent multi-lateration. While this works well under line-of-sight (LOS) propagation conditions, the performance of these approaches deteriorates heavily in indoor environments where multipath propagation is predominant. This is why approaches for dealing with multipath (MP) propagation have been investigated in recent years~\cite{AdityaMPCintro, DaradariIndoor}. Specifically, the large bandwidths offered by modern RF positioning systems such as ultra-wideband (UWB) allows for an increased temporal (and therefore spatial) resolution of propagation paths \cite{MolischUWBIEEEProc}.

Since MP propagation produces valuable spatial information~\cite{LeitingerPINFMP, Meissner2014} it is desirable to exploit it for positioning. However, the complete extraction of such spatial information embedded in channel measurement (CM) vectors is still an active research area, as different propagation effects such as reflection, diffraction or scattering contribute different kinds of information, some of which are hard to model analytically. 

Methods such as channel simultaneous localization and mapping (C-SLAM)~\cite{LeitingerSLAMBP, GentnerChannelSlam3}, error mitigation (EMI)~\cite{WymeRangingError, stahli1337, fontaine_edge_2020}, and fingerprinting (FP)~\cite{niitsoo2019deep, KramCIRFP, WangCSI2017, TsengCIRFP2018} exploit or mitigate the spatial characteristics of environment-related signal propagation using information from CM. In contrast to C-SLAM and EMI, which both typically rely on observation likelihood modeling, FP formulates a regression or classification task with either the complete CM or extracted features thereof as the input, and the positions as an output of a neural network. Still, unlike C-SLAM and EMI, FP exploits additional, diffuse multipath information and hence, can perform well in densely cluttered environments, see Fig.~\ref{fig:propconds}: the LOS component that EMI mainly relies on, and strong MPCs caused by reflecting surfaces that C-SLAM requires, are not as readily available.

However, state-of-the-art FP has a number of downsides. 
First, as it directly estimates the position it does not model an observation likelihood explicitly. 
This leaves much of its potential unused as information fusion with other sources of spatial information such as time-of-flight (TOF) is only possible on the state level of the tracking target. 
Thus, to fuse all these approaches on the observation level, we need to represent spatial information as an observation likelihood model. 
Second, FP requires large, spatially dense datasets.
This is difficult in practice as such datasets must be acquired with expensive positioning reference systems and adapted to environment changes. 
Third, FP only outperforms EMI in areas with complex signal propagation and low LOS, i.e., in smaller areas within the environment (such as densely packed production spaces).
Thus, it is desirable to resort to FP only in parts of the environment to lower the database acquisition and maintenance effort. 
Fourth, FP requires all CMs at a central point for processing which results in significant communication load. Hence, in practice FP requires compressed CMs to scale to a large number of devices.

To address these issues jointly (i.e., FP observation likelihood modeling, spatially sparse datasets, and compact CM representation), we propose Gaussian process regression (GPR) of features extracted from CM using an autoencoder (AE) neural network \cite{CIRPRESSOR} and model-based propagation-related features \cite{WymeRangingError, KramCIRFP, HuangTransFeatures} to generate a statistical observation likelihood model. 
Unlike state-of-the-art FP methods based on deep (convolutional) neural networks (CNNs) \cite{n2018enhancement, niitsoo2019deep, TsengCIRFP2018}, GPR requires less labeled data. 
Furthermore, it estimates an observation likelihood model based on proximity to FP observations instead of positions directly. Thus, it can be directly integrated into a tracking filter, e.g., a particle filter (PF).

We study different architectures and parameter configurations for both AE and GPR using real-world data obtained in a realistic industrial indoor environment. We benchmark the positioning and sensor fusion capabilities against state-of-the-art EMI and FP approaches in terms of positioning accuracy and data requirements. Our evaluation focuses on a realistic indoor industrial environment with various interfering objects and a low number of only three anchors. 
Our position estimates are more accurate than those of EMI for a dense FP dataset, while state-of-the-art FP based on a CNN (CNN-FP) only slightly outperforms our model (as expected). However, when considering a spatially sparse FP dataset, unlike CNN-FP, our model estimates its own reliability and therefore exhibits smaller performance decrease, resulting in a significantly more accurate tracking than the baselines. Therefore, accurate tracking is possible with smaller databases that are easier to obtain and maintain in practical applications.

The remainder of the article is structured as follows. Sec.~\ref{sec:system_model} describes the system model and formulates the problem. Sec.~\ref{sec:rw} discusses related work. Secs.~\ref{sec:infex} and \ref{sec:obs} describe our solution. Sec.~\ref{sec:evaluation} describes the evaluation setup. Sec.~\ref{sec:featan} evaluates the proposed feature selection, while ~\ref{sec:posres} discusses experimental results. Sec.~\ref{sec:conclusion} concludes.

\section{Multipath-assisted Agent Tracking in UWB channels}
\label{sec:system_model}

\begin{figure}[t!]%
    \centering%
    \hspace{-6mm}\includegraphics[width=\linewidth]{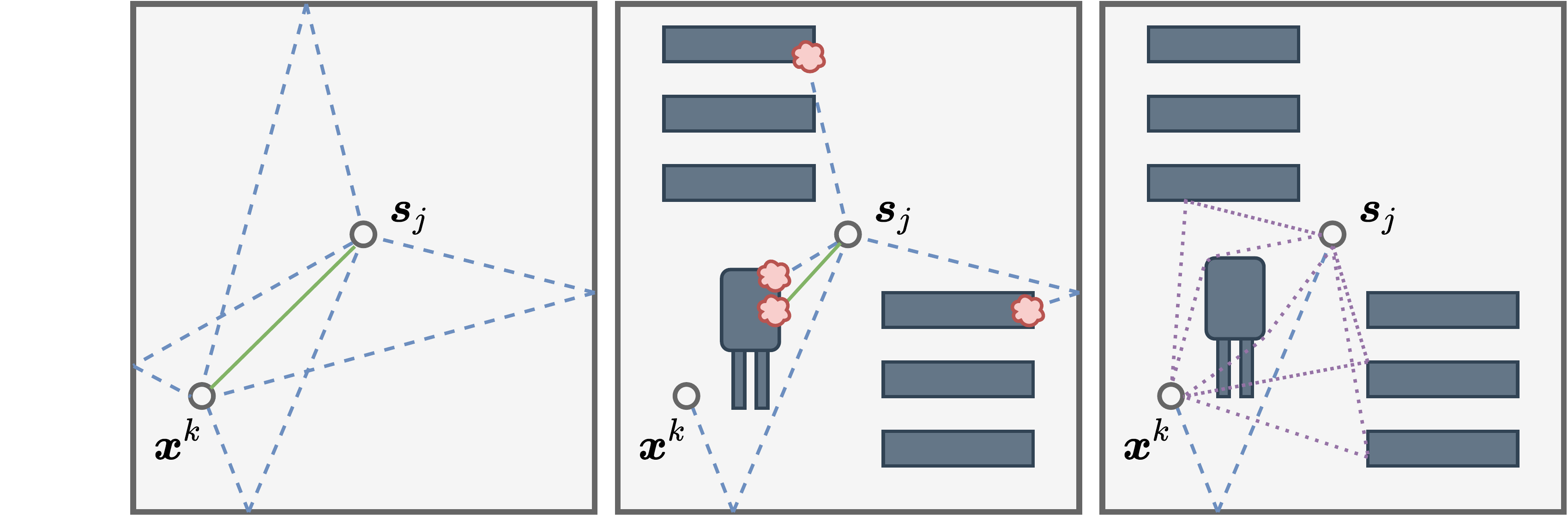}%
    \caption{Propagation conditions in empty (left) and cluttered (middle/right) rooms: In empty rooms we see clear LOS connections (green straight line) between $\boldsymbol{x}^k$ and $\boldsymbol{s}_{j}$ but walls cause distinguishable, deterministic specular reflections (blue dashed lines). 
    When multiple objects are introduced to the environment, these deterministic propagation components are shadowed (red clouds, center), while the objects themselves cause diffuse interaction like diffraction and scattering (purple dotted line, right).}%
    \label{fig:propconds}%
\end{figure}

The spatial information contained in CMs can be modelled as follows (see Fig.~\ref{fig:propconds}): 
At timestep $k$, an anchor, located at $\boldsymbol{s}_{j}$, sends a signal $s(t)$ of pulse length $T_p$ in the UWB baseband to a mobile agent at $\boldsymbol{x}^k$. 
The agent node receives
\begin{equation}
    r(t) = s(t) * h(t) + w(t),
\end{equation}
where $s(t) * h(t)$ represents the complex spatial component (via a convolution of $s(t)$ with the channel impulse response $h(t)$) and $w(t)$ describes the complex non-spatial noise components, consisting of, e.g., thermal noise or errors introduced by the processing chain. 
$w(t)$ is typically modelled as a temporally uncorrelated white Gaussian noise $w(t) \sim \mathcal{CN}(0, \sigma_{w}^2)$ of the double-sided power-spectral density of $N_{0}/2$~\cite{LeitingerPINFMP}.

$h(t)$ contains both spatially deterministic and diffuse components. 
The first corresponds to distinguishable propagation paths. 
The latter contains unsolvable multipath components (MPCs). 
Depending on the existence $\epsilon_{0} \in \{0,1\}$ of the LOS component, each of the $M$ or $M+1$ propagation paths can be defined by a delay $\tau_{m'}$ and corresponding signal attenuation $\alpha_{m'}$. 
The diffuse noise component $\nu(t)$ contains additional MPCs caused by the environment through diffraction, rough scatterers or higher order reflections. 
This process is usually modeled as a zero-mean random process of the time-dependent power delay profile $S_{\nu}(\tau) \delta(\tau-u) = \mathbb{E}\{\nu(t) * \nu(u)^* \}$\cite{Meissner2014}:
\begin{equation}
        h(t) = \underbrace{ \epsilon_{0} \alpha_{0} \delta(t-\tau_{0}) + \sum_{m=1}^{M} \alpha_{m}\delta(t-\tau_{m})}_{\mathrm{deterministic}} + \underbrace{\nu(t)}_{\mathrm{diffuse}}.
\label{eq:chmod1}
\end{equation}

To employ this information to track mobile agents, at each timestep $k$, a set of CMs $\mathcal{R}^k = (r_{1}^k(t),..., r_{j}^k(t), ...r_{J}^k(t))$ is available from anchors located at $\boldsymbol{x}_j$, each consisting of a complex time series of length $L$, that can alternatively be interpreted as a $L \times 1$ complex vector $\boldsymbol{r}_{j}^k$. 
The goal of a tracking algorithm is then to use subsequent sets of CMs and initial agent position $ \boldsymbol{x}^0$ to track the agent position $\boldsymbol{x}^k$:
\begin{equation}
    \hat{\boldsymbol{x}}^k = \underset{\boldsymbol{x}^k}{\operatorname{argmax}} \;  p(\boldsymbol{x}^k| \mathcal{R}^k,  \mathcal{R}^{k-1}, ..., \mathcal{R}^{0}, \boldsymbol{x}^0).
\end{equation}

Classical approaches~\cite{Zafari} interpret the maximum of the CM as corresponding to the direct LOS connection. 
Thus, each $\boldsymbol{r}_j^k \in \mathcal{R}^k$ is compressed into a single value, i.e., a noisy estimate of $\hat{d}_{j}^k = ||\boldsymbol{x}^k - \boldsymbol{s}_j||_{2}$ or a similar quantity. 
However, not only is there much more information contained in the channel as introduced with \eqref{eq:chmod1}, in indoor environments the LOS assumption is also often violated~\cite{KramCIRFP, TALLA}. 
It is therefore desirable to exploit more spatial information from $\mathcal{R}^k$ if necessary, while including (applicable) LOS information.

\section{Related Work}
\label{sec:rw}

\begin{figure*}[tb]
    \centering
    \includegraphics[width = 0.7\linewidth]{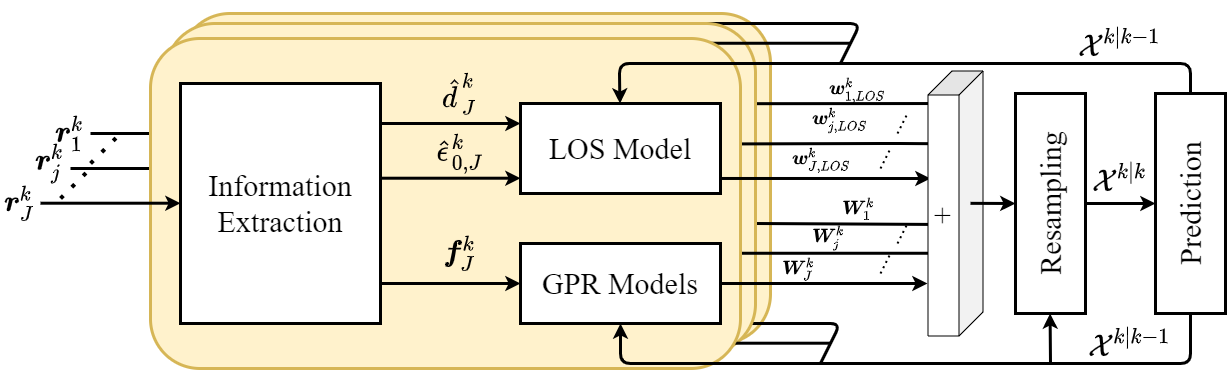}
    \caption{The processing pipeline for the proposed tracking method: 
    First information is extracted from the CMs $\boldsymbol{r}$ to obtain $\hat{d}, \hat{\epsilon}_{0}$ and $\boldsymbol{f}$ (Sec.~\ref{sec:infex}). 
    Then, the observation likelihoods are calculated for the LOS component and feature vector in the form of log-likelihood weights (Sec.~\ref{subsec:obsmod}) for predicted particles $\mathcal{X}^{k|k-1}$. 
    The weights from the different information sources are then summed up and used for resampling, resulting in the updated particles $\mathcal{X}^{k|k}$.}
    \label{fig:sys_over}
\end{figure*}

A variety of different methods for exploiting additional information in CMs have been published. 
For instance, error mitigation (EMI) approaches rely on the correction of ranges by estimating systematic offset statistics with regression~\cite{WymeRangingError, stahli1337} or classify adverse channel conditions~\cite{CUILOSNLOSWAVELET2021, WuLOSNLOSSVM, StahliNLOS}. 
Hence, they produce a correction indicator that can be used for the enhancement of LOS range estimates by offset mitigation, error statistic estimation, or the exclusion of unreliable LOS ranges. 
Although EMI leads to a significant performance accuracy gain over classical methods, it only focuses on the LOS component. Hence, it requires anchor redundancy and does not consider additional spatial components as a separate source of spatial information.

To circumvent these limitations, C-SLAM \cite{GentnerChannelSlam3, LeitingerSLAMBP, Kasarekt, XuhongCSLAM} relies on exploiting MPCs rather than mitigating their effect on range estimation. 
Each CM is first decomposed into a set of MPC delays (or, additionally amplitudes~\cite{LeitingerMagnitude}), each of which is then associated with a specific reflecting surface or strong scatterer within the environment to produce additional delay measurements originating from a corresponding virtual anchor. 
This allows for a reliable and accurate tracking even with a limited number of anchors. However, these approaches rely on subsequent observation of strong MPCs and exhibit problems in densely packed spaces such as industrial environments, where smaller objects create diffuse MPCs and obstruct deterministic components which hinder the tracking of virtual anchor hypotheses due to visibility constraints.

Instead, FP methods rely on sets of previously recorded positions and CM sets (i.e., fingerprints / FPs). 
The position estimate is then obtained by relating new observations to the FPs, potentially combining the estimate with a hand-engineered~\cite{Niitsoo2018} or learned~\cite{Feigl2018RNN} dynamics model. 
RF-FP is a topic where a huge variety of methods exists especially for signals of lower complexity such as Wi-Fi or Bluetooth RSS~\cite{joquimmeta}. 
However, due to the complexity of CMs, FP has mostly been realized via CNNs that solve a regression problem by approximating a function that matches the training data~\cite{niitsoo2019deep, TsengCIRFP2018, WifiDLFP}. 
Unlike EMI and CSLAM, FP also exploits diffuse MPCs implicitly as it considers the spatial characteristics of the complete CMs. 
State-of-the-art FP methods directly estimate positions so that a fusion with other information sources is only possible on the state level, hindering a combination with other information contained in the CM, like TOFs. Existing FP methods yield high positioning accuracy but they need large, spatially dense datasets of complete CMs which are hard to obtain and maintain in practice. It is hence desirable to resort to spatially sparse data and a compact CM representation.

\textbf{Contribution.} We propose to use smaller and spatially sparse datasets to train a GPR~\cite{Rasmussen}, which implies observation likelihood models by modeling spatial distributions of the outcomes of different features extracted from CMs. 
Previous work proposed to compress CMs using AE neural networks~\cite{CIRPRESSOR}, resulting in a compact representation of CMs. 
Furthermore, various informative features of CMs have been deducted for positioning-related tasks~\cite{WymeRangingError, HuangTransFeatures, KramCIRFP}.

We use these GPs to combine the spatial information with separately extracted LOS TOFs as an observation model in a particle filter (PF), fusing the obtained spatial information on the state level. 
GPRs have been successfully employed for positioning with complex, environment-related signals such as magnetic field measurements~\cite{kokgpr, sieblermagnetic} and also for RSS measurements~\cite{VeldeCoop, KramUKR, MiyagusukuPLGPR}. 
Since we use CM, however, we apply an additional information extraction layer (see Sec.~\ref{sec:infex}), evaluating different feature combinations to represent the spatial information contained in CMs. 
GPR has also been used in relation to CMs~\cite{nguyengprmpc} to estimate deterministic MPCs for C-SLAM. Instead, we employ GPR as a tool for observation modeling in FP (see Sec.~\ref{sec:obs}).

Our method differs from the state-of-the-art in CM-assisted positioning. 
First, unlike EMI and LOS approaches, we use additional spatial information contained in MPCs for tracking. 
Therefore, we can track with lower numbers of anchors and in environments without LOS. 
Second, unlike CNN-FP, we use a compact feature-space representation of the CM,  reducing communication bandwidth requirements.
Furthermore, as we use GPR we learn observation likelihood models that also estimate the information reliability. 
We can thus combine the information on an observation level, providing reliability estimates for all information sources and so, do not require spatially complete data.

\section{Information Extraction}
\label{sec:infex}

We extract spatially significant signal features from each complex-valued CM vector $\boldsymbol{r} \in \mathbb{C}^{L}$, resulting in a low-dimensional representation $\boldsymbol{f} \in \mathbb{R}^{F}$, see \textit{Information Extraction} in Fig.~\ref{fig:sys_over}. 
We omit anchor and time indices for clarity (e.g., $\boldsymbol{r} \triangleq \boldsymbol{r}_{j}^{k}$) if it is clear from the context.

\subsection{LOS Extraction}
\label{subsec:LOS}

First, we classify if $\boldsymbol{r}$ contains a LOS component, using a state-of-the-art LOS identification algorithm \cite{stahli1337}. Hereby, we calculate a LOS quality parameter $\beta_{0}$ using an anomaly detection algorithm, trained on a set of \textit{unlabeled and anchor- and environment-independent} CMs obtained in a pure LOS environment. Therefore, the algorithm provides an estimate $\hat{\epsilon}_{0} \in \{0,1\}$ of the LOS existence $\epsilon_{0}$. 
If $\beta_{0}$ is above a threshold, we assume that $\hat{\epsilon}_{0}=1$. 
In this case, we obtain and evaluate a distance estimate $\hat{d}$.

\subsection{Feature Extraction}
\label{subsec:infex:feature_extraction}

We select well-known propagation-inspired features: the signal energy (ENG), signal decay times (SDT50/SDT75), minimum delay index (MDI), root-mean-square delay spread (RMSDS), Ricean k-factor (RKF), skewness (SKE), and kurtosis (KUR). 
For details on their computation and additional features we refer the interested reader to~\cite{WymeRangingError, HuangTransFeatures, KramCIRFP}. 
We compare the suitability of the selected features in Sec.~\ref{sec:exp:feature_selection}.

However, previous work~\cite{CIRPRESSOR} has shown that the latent space representation of CMs learned via autoencoders (AEs) compactly represents the information contained in CMs. Hence, we investigate AE-based feature extraction for our positioning task. 
We train the encoder and the decoder by processing the CM magnitudes $|\boldsymbol{r}|$. 
The AE implicitly yields a latent space representation $\boldsymbol{a} \in \mathbb{R}^{A}$. 
As AEs work unsupervised, we train our AE on an \textit{unlabeled} set of training data obtained in the same environment. 
Hence, the AE is trained to compress the information from CM without considering the observation position, i.e., there is no positional bias. 
As it is unclear which (non-)spatial signal-components are encoded in the latent space and which of them are beneficial for tracking, we analyze different subsets of the elements of $\boldsymbol{a}$ in Sec.~\ref{sec:evaluation}.

By combining the extracted features, we obtain a feature vector $\boldsymbol{f} \in \mathbb{R}^F$ containing elements of the propagation $\boldsymbol{p}$ and/or AE features $\boldsymbol{a}$ in addition to the LOS distance $\hat{d}$ (if $\hat{\epsilon}_0 = 1$). 
This is depicted in Fig.~\ref{fig:sys_over}.

\section{Information Fusion}
\label{sec:obs}

To exploit the extracted features $\boldsymbol{f}$ for positioning, however, they have to be combined with previously collected observations (i.e., FPs) as a straightforward geometrical modeling is impossible. 
Specifically, in this case a sparse spatial representation is of interest: 
In cases where a LOS connection to anchors is available, $\hat{d}_j^k$ provides enough spatial information for accurate tracking. 
So, there is no need to assist tracking with recorded observations. 
However, in cases where (especially diffuse) MPCs dominate propagation, the LOS component is either not available or superimposed by MPCs, so that $\boldsymbol{f}$ has to provide additional spatial information to maintain accurate tracking. 

\subsection{Tracking Problem}

Based on consecutive feature and range estimates extracted from sets of observed CMs $\mathcal{R}^k$, we describe our estimation problem of tracking the agent state $\boldsymbol{x}^k$. 
An FP database $\mathcal{Y}$ and an a-priori state estimate $p(x^k |x^{k-1})$ provide additional spatial information:

\begin{equation}
    p(\boldsymbol{x}^{k}| \mathcal{R}^{k}, \mathcal{Y}) \propto p(\mathcal{R}^{k}| \boldsymbol{x}^k , \mathcal{Y}) \; p(x^k |x^{k-1}). 
    \label{eq:trackprob}
\end{equation}

Using the extracted feature vectors $\boldsymbol{f}^k_j$ (Sec.~\ref{sec:infex}), the observation likelihood $p(\mathcal{R}^{k}| \boldsymbol{x}^k, \mathcal{Y})$ can be reformulated as a combination of the likelihoods of the extracted LOS ranges and features. 
While the observation assumption of the LOS features depends on $\hat{\epsilon}_{0}$, the observation likelihood is given by a simple geometric model and additive Gaussian noise with standard deviation $\sigma_d$, which is either available from device documentation or can be estimated. 
Hence, $p(\hat{d}_j^k |\boldsymbol{x}^k) \sim \mathcal{N} (||\boldsymbol{x}^k - \boldsymbol{s}_{j}||, \sigma_d^2)$. 
The feature observation likelihoods $p(z_{j,f}^{k}|\boldsymbol{x}^k, \mathcal{Z}^{j,f})$ are inferred by training data $\mathcal{Z}^{j,f}$ for each feature/anchor combination, so that we factorize $p(\mathcal{R}^{k}| \boldsymbol{x}^k , \mathcal{Y})$ from \eqref{eq:trackprob} into
\begin{equation}
    p(\mathcal{R}^{k}| \boldsymbol{x}^k, \mathcal{Y}) \propto \underbrace{\prod_{j=1}^J p(\hat{d}_j^{k}|\boldsymbol{x}^k)^{\hat{\epsilon}_{0}}}_{\mathrm{LOS}} \times \underbrace{\prod_{j=1}^J \prod_{f=1}^F p(z_{j,f}^{k}|\boldsymbol{x}^k, \mathcal{Z}^{j,f})}_{\mathrm{Features}}
    \label{eq:featslikel}
\end{equation}
to separate the contributions from each feature. Note that hereby we assume statistical independence between the features.

\subsection{Feature Observation Likelihood Representation}
\label{subsec:obsmod}

Unlike it is the case for $\hat{d}_{j}^k$, for the extracted feature vector a straightforward geometrical derivation of the observation likelihood is not possible: 
It contains information on environment-specific propagation conditions, both spatially deterministic and diffuse. 
Therefore, we apply GPR~\cite{Rasmussen} to estimate the observation likelihood from our FP database $\mathcal{Y}$. 
We train a GPR model $g_{j,f}$ for each feature/anchor combination using $Z_{f,j}$ training data $\mathcal{Z}^{j,f} = \{(z_{j,f}^1, \boldsymbol{x}_{j,f}^1), ..., (z_{j,f}^{Z_{f,j}}, \boldsymbol{x}_{j,f}^{Z_{f,j}})\}$, consisting of positions $\boldsymbol{x}_{j,f}$ and feature observations $z_{j,f}$. 
We fit an individual standard scaler on all $\mathcal{Z}^{j,f}$ for each feature/anchor combination for numerical stability and apply it to all FPs and test data.

The GPR models the statistics of observed data by fitting covariance kernels $k_{j,f}(\boldsymbol{x}_{1}, \boldsymbol{x}_{2})$. 
They model the statistical dependence between observations based on their positions in space. 
If the process is intrinsically stationary, the covariance between observations only depends on the Euclidean distance between them. 
Thus, it can be modeled using a distance-dependent stationary kernel function $k_{j,f}(||\boldsymbol{x}_{1}- \boldsymbol{x}_{2}||)$. 
$k_{j,f}$ is obtained by fitting a set of hyperparameters to the training data. 
As we apply GPR to the extracted features, a straightforward derivation of a suitable kernel function is impossible. 
Hence, we evaluate different kernel function used in related work in Sec.~\ref{sec:evaluation}. 
Each trained GP then regresses the observation likelihood function $p(z_{j,f}^{k}|\boldsymbol{x}^k, \mathcal{Z}^{j,f})$, represented by a Gaussian distribution:
\begin{equation}
    p(z_{j,f}^{k}|\boldsymbol{x}^k, \mathcal{Z}^{j,f}) \approx g_{j,f}(\boldsymbol{x}^{k}) \sim \mathcal{N}(\mu_{f,j}(\boldsymbol{x}^{k}), \sigma_{f,j}^2(\boldsymbol{x}^{k})).
    \label{eq:log_likelihood}
\end{equation}

\begin{figure}[t!]
    \centering
    \includegraphics[height=4.0cm, trim= 20 0 95 0, clip]{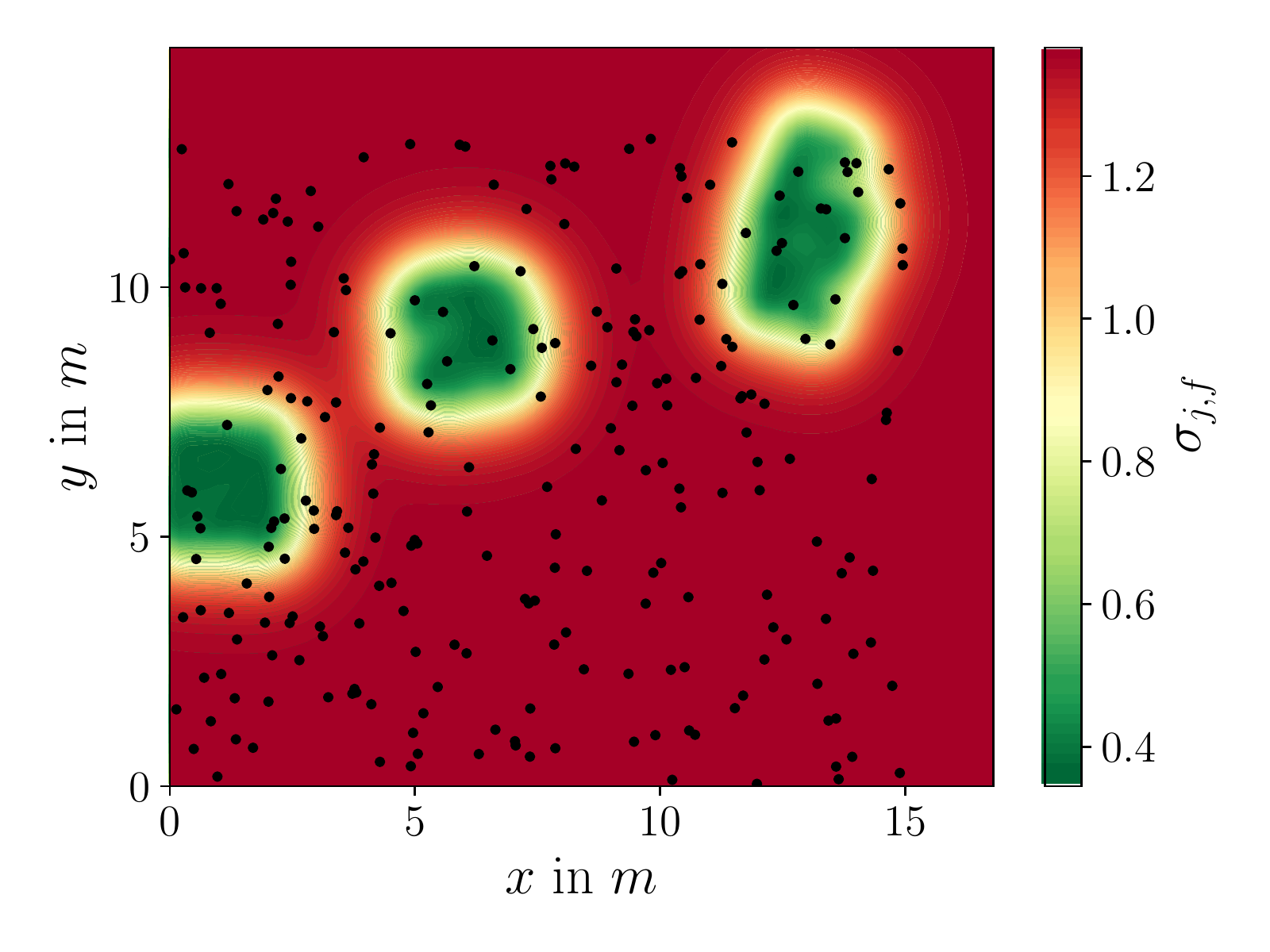}
    \includegraphics[height=4.0cm, trim= 53 0 10 0, clip]{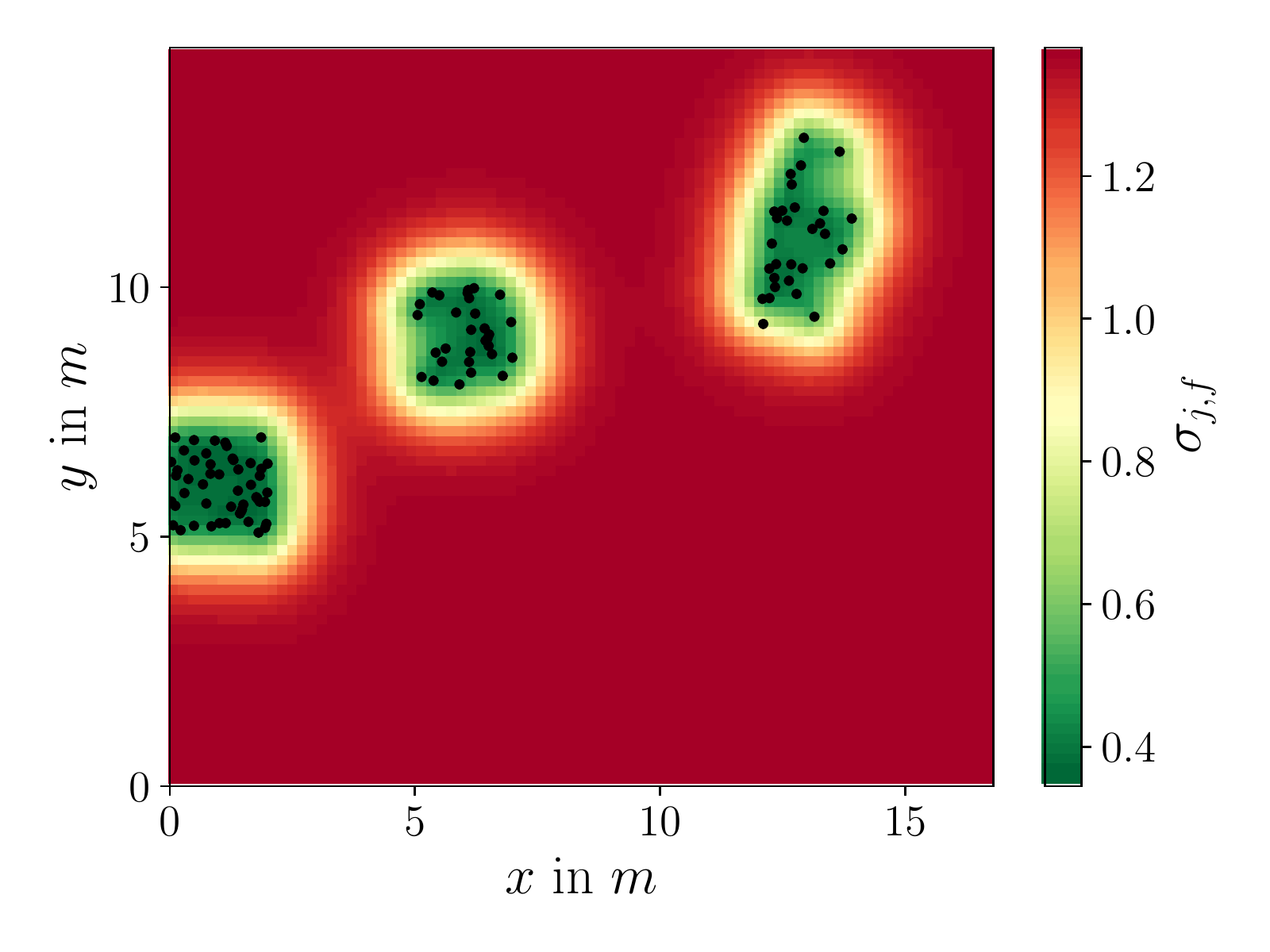}
    \caption{Exemplary spatial distribution of $\sigma_{j,f}$ for a spatially dense (left) and sparse (right) dataset, each in arbitrary units due to scaling. 
    The positions of the data recordings $\mathcal{Z}^{j,f}$ are indicated as black dots.}
  \label{fig:spatdist1} 
\end{figure}

The kernel depends on the distance between $\boldsymbol{x}^k$ and each previous observation position $\boldsymbol{x}_{j,f}$. 
Hence, $\sigma_{f,j}$ increases significantly in areas within the environment where no information is available from data recorded in close proximity, while $\mu_{f,j}$ converges to the global mean. 
This is depicted in Fig.~\ref{fig:spatdist1}: 
Each subfigure visualizes $g_{j,f}(\boldsymbol{x}^{k})$ for a single, arbitrary feature value. 
On the left side, $\mathcal{Z}^{j,f}$ includes recordings distributed over most of the environment. 
Therefore, $\sigma_{f,j}$ is low in most of the area, only increasing toward the outer edges of the environment. 
On the right side, we see a distribution based on a smaller, sparse set of data, which, for the application case could be consisting of only data in close proximity to the objects, where diffuse multipath is dominant. 
Consequently, $\sigma_{f,j}$ increases rapidly (red color) outside of these areas, indicating a decrease in confidence that can be reflected in the weighting of the information source. 
This is beneficial for positioning, as in these areas, without interfering objects, the presence of a usable LOS component is more likely, so that $\hat{d}_j$ can serve as a primary source of spatial information.

\subsection{Tracking Filter}
\label{subsec:trackfilter}
To now solve the tracking problem formulated in .~\eqref{eq:trackprob}, we set up a PF~\cite{PartiGenerisch, NIAParti}, visualized on the right side of Fig.~\ref{fig:sys_over}. 
We represent $p(\boldsymbol{x}^k)$ by $P$ agent particles $X^k_p \in \mathcal{X}^k; p \in\{1,...,P\}$ with an agent state vector of position and velocity hypothesis and weights, i.e., $X^k_p =\{[\boldsymbol{x}_p^k ; \boldsymbol{v}_{p}^k], w_p^k\}$. 
In the prediction step, we propagate the set of particles $\mathcal{X}^k$ using constant velocity model fitted to the agent state dynamics to realize $p(x^k |x^{k-1})$, yielding the set of predicted particles $\mathcal{X}^{k|k-1}$.

For the update step, we first use our information extraction pipeline from Sec.~\ref{sec:infex} to obtain $\hat{\epsilon}_{0, j}^k$, $\hat{d}_{j}^{k}$ and $\boldsymbol{f}_{j}^{k}$. 
Next, we obtain the weights of each individual particle by evaluating our models for all observation log-likelihoods: 
In the LOS case ($\hat{\epsilon}_{0} = 1$), the weight $ w_{j,LOS,p}^k$ of the obtained TOF is given by the model for $p(d_j^{k}|\boldsymbol{x}^k)$:
\begin{equation}
    w_{j,LOS,p}^k = \left(-\frac{\left(\hat{d}_{j}^k - \left\Vert\boldsymbol{x}_p^{k|k-1} - \boldsymbol{s}_{j}\right\Vert\right)^2} {2 \sigma_d^2} - \log\left(\sqrt{2 \pi} \sigma_d\right)\right)^{\hat{\epsilon}_{0,j}^{k}}.
\end{equation}
Hence, for the NLOS case ($\hat{\epsilon}_{0} = 0$) we assign uniform weights. 
Then, for all selected features, the corresponding weight $w^k_{j, f, p}$ is given by the observation log-likelihood of the GP modeling $p(z_{j,f}^{k}|\boldsymbol{x}^k, \mathcal{Z}^{j,f})$:
\begin{equation}
    w^k_{j, f, p} = - \frac{\left(z_{j, f}^{k} - \mu_{j, f} (\boldsymbol{x}^{k|k-1})\right)^2}{ 2 \left(\sigma_{j, f}(\boldsymbol{x}^{k|k-1})\right)^2} - \log\left(\sqrt{2 \pi} \sigma_{j, f}(\boldsymbol{x}^{k|k-1})\right).
\label{eq:loglk}
\end{equation}
For each anchor, we assign $\boldsymbol{W}_{j}^{k} = \left[\boldsymbol{w}^k_{j, 1}, ..., {w}^k_{j, F}\right]$.
We then normalize all weights for numerical stability so that $\sum_{p=1}^{P} \exp(w^k_{j, f, p}) = \sum_{p=1}^{P} \exp(w_{j,LOS,p}^k) = 1 \forall j,f$. 
Finally, we sum up all the obtained weights to model the combined likelihood $p(\mathcal{R}^{k}| \boldsymbol{x}^k, \mathcal{Y})$:
\begin{equation}
    w^{k}_{p} = \sum_{j=1}^J \left( \hat{\epsilon}_{0} w^k_{j,LOS,p} + \sum_{f=1}^F w^k_{j,f,p} \right).
\end{equation}
We then apply sequential importance resampling~\cite{PartiGenerisch} on the particles $\tilde{\mathcal{X}}^{k|k}$ with the updated weights $w^{k}_{p}$, resulting in the updated set of agent particles $\mathcal{X}^{k|k}$ modeling the final estimate $p(\boldsymbol{x}^{k}| \mathcal{R}^{k}, \mathcal{Y})$. 
We then obtain the position estimate by calculating the mean of all particle positions $\boldsymbol{x}_p^{k|k}$.

\section{Evaluation}
\label{sec:evaluation}

Before we evaluate the proposed method in Secs. \ref{sec:featan} and \ref{sec:posres}, we introduce the measurement setup, parameter configurations, and baseline methods.

\subsection{Dataset}

\begin{figure}[t!]
    \centering
    \includegraphics[width=0.9\linewidth, trim=0 700 100 100, clip]{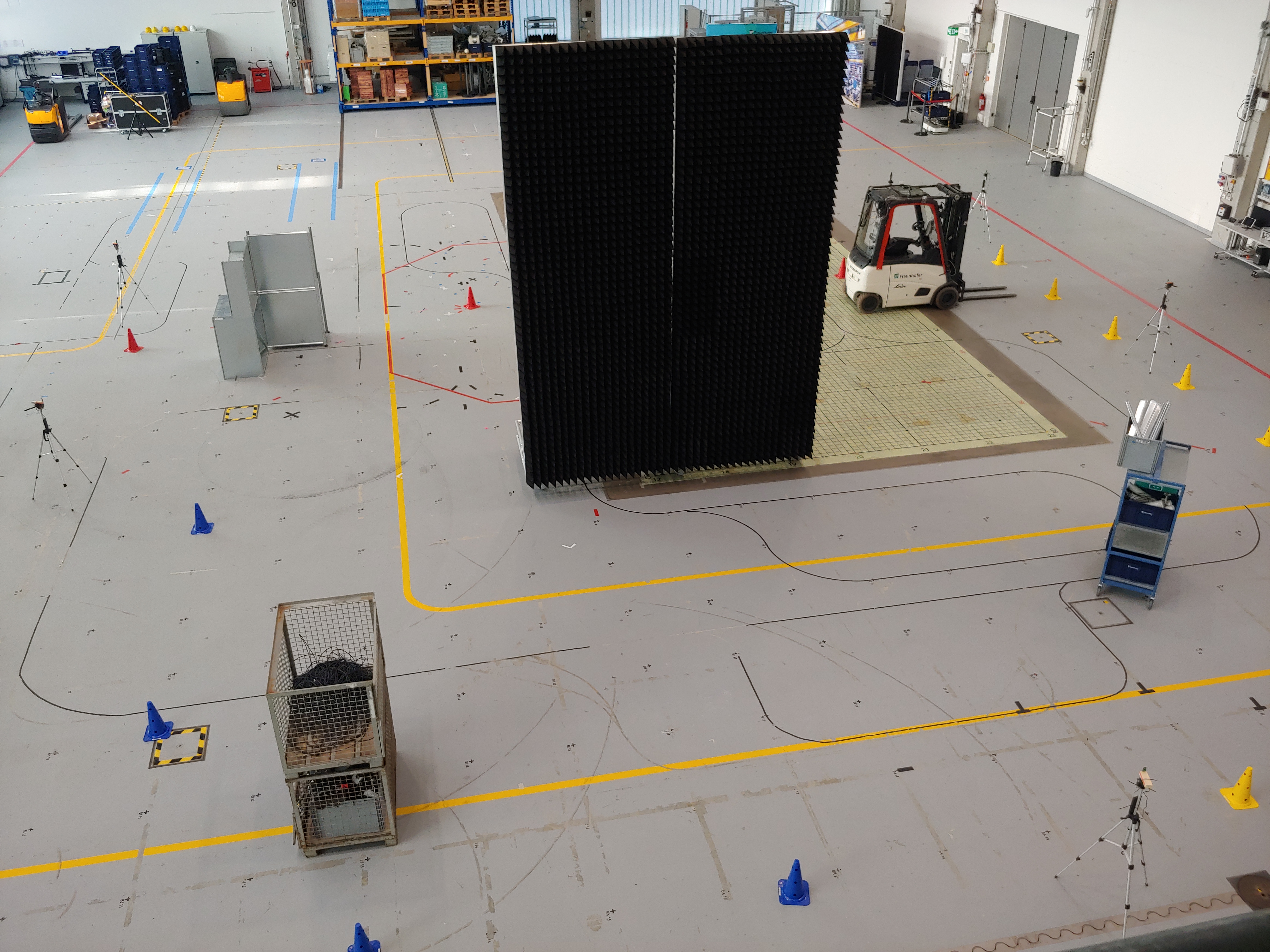}
    \caption{Our measurement environment.}
    \label{fig:meas_photo}
\end{figure}

For evaluation, we use a dataset recorded in a scenario that models an industrial environment, see Fig.~\ref{fig:meas_photo}, which includes various metallic objects (i.e., absorber/reflector walls, metal shelves, and an industrial vehicle). 
The positions of theses objects are indicated as black boxes (see Fig.~\ref{fig:emi_traj3_anchs}) or grey boxes (see Fig.~\ref{fig:spatkurt1}). 
The environment is located in a large industrial hall with various additional objects and reflecting concrete walls, floors, and ceilings. 
For the evaluation, we use three receiving anchors located around the recording area (see the black triangles in Fig.~\ref{fig:emi_traj3_anchs}). 
We use the Decawave DW1000 \cite{KulmerDecawave} to obtain UWB CMs at a bandwidth of $499.2$\si{MHz} and center frequencies of $4-6$ \si{GHz} and a QualiSys optical reference system with a accuracy in the \si{mm} range for position reference data.

Our full dataset recording is depicted in Fig.~\ref{fig:spatkurt1} (see the black dots), covers most of the area, and consists of approx. $16{\small,}000$ datapoints. 
This dataset is ideally for FP methods, as representative data are available in almost all positions reached in the positioning task. 
For training we investigate the effects of both a spatially \textit{sparse} and \textit{full} dataset. 
The sparse dataset is much smaller ($1{\small,}600$ datapoints) as we only use a subset of the recorded data close (distance $< 1.5$\si{m}) to the objects causing MPCs and shadowing of the LOS component, as indicated as black and white dots in Fig.~\ref{fig:spatkurt2}. 
Hence, this dataset only contains data from areas where classical or EMI approaches do not observe sufficient information for tracking. 

\subsection{Architectures, Parametrization and Optimization}

\textbf{GPR}: We use the gpytorch Python library~\cite{Gardner2018GPyTorchBM} to train GPR. 
We compare commonly used RBF, Matern52, and Matern32 kernels, all of which are commonly used stationary kernels~\cite{Rasmussen}.
In a preliminary study we obtained the best results with Matern52, which relies on modified Bessel functions that fit well to 2D wave propagation.
Hence, we only show results with this kernel. 
We split data into training ($90 \%$) and validation ($10 \%$) sets and train the GPs for 500 iterations, storing the model with the lowest sum of log-likelihood distances (see ~\eqref{eq:log_likelihood}) on the validation data.

\textbf{AE}: We investigate three different AE architectures, each with $\{4, 6, 8\}$ latent variables. 
We train them on \textit{unlabeled} CM magnitudes $|\boldsymbol{r}|$ from the dataset. Hereby, we use CMs from more anchors resulting in approx. $37{,}000$ datapoints, split into $80 \%$ training and $20 \%$ validation data. 
Our first architecture (FCN-AE) employs a fully connected AE with two hidden layers of size 150 and 80 per encoder and decoder and a rectified linear unit (ReLU) activation function. 
Our second architecture (CNN-AE-S) uses the best-performing AE from~\cite{CIRPRESSOR} with three convolutional layers per encoder/decoder with exponential linear unit (ELU) and ReLU activation functions. 
Our third architecture (CNN-AE-P) is a modification of CNN-AE-S that employs the same layer sizes throughout the whole model, such that we apply a max pooling of size 2 after the activation function of each convolutional layer in the encoder. This halves the layer size. 
In the decoder, we replace the convolutional layers with transposed convolutional layers (up-convolutions), which symmetrically upsample and double the layer size. 
The idea is to reduce the number of parameters and thus, to improve generalization and to increase the receptive field of the convolutional layers.

\textbf{PF}: We evaluate the PF derived in Sec. \ref{subsec:trackfilter} with $P = 10{\small,}000$ particles and repeat all experiments twice for feature selection and 200 times for the final positioning results to assure stability of our results. 
We initialize the original state estimate by sampling from $\mathcal{N}(\boldsymbol{I}_{2})$. 
We initialize all $\boldsymbol{x}_{p}^{0}$ by sampling from $\mathcal{N}(\boldsymbol{x}_{0, \text{ref}}, 1 \si{m}^2 \boldsymbol{I}_{2})$, where $\boldsymbol{x}_{0, \text{ref}}$ is the initial position in the reference data. 
For $\boldsymbol{v}_{p}^{0}$, we initialize the $x$ and $y$ components by sampling from $\mathcal{N}(\mu_{v}, \sigma_v^2)$, where $\mu_{v}$ and $\sigma_v$ are obtained by taking the absolute mean and standard deviation of the velocities in our training data.

\subsection{Baseline Methods}

To evaluate our tracking method against state-of-the-art, we compare the results with both EMI and CNN-FP approaches:

\textbf{EMI}. We first detect NLOS~\cite{stahli1337} to exclude unreliable TOFs using a LOS quality indicator threshold ($\beta$ of $1.5$) obtained by analysis of the data.
We combine the TOFs classified as LOS in a range-based PF. For a fair comparison, we implement the same observation likelihood and dynamics of our pipeline. 
We consider this a feasible state-of-the-art approach, however there are also other methods for error mitigation, including body shadowing models~\cite{coenebody} and TOF regression~\cite{FeiglML4CIR2Toa}.

\textbf{CNN-FP.} We train a CNN-based position regressor~\cite{niitsoo2019deep}, on the same data we use to train our GPR models. 
We use a $6$ layer CNN architecture that consists of $4$ convolutional layers for feature extraction and $4$ fully connected layers for position estimation. 
The input tensor consists of the recorded complex CMs shifted by the estimated TOFs. 
We use ReLU activation for all layers, except for the last layer that uses a linear activation. 
We omit local pooling as this may harm the accuracy on time-series tasks~\cite{fawaz2019deep}. 
Nevertheless, to enhance the receptive field of the CNN we enlarge the kernel sizes with the depth of the model: 
The first convolutional layer uses a kernel size of ${[1 \times 3]}$, the second ${[2 \times 5]}$, the third ${[3 \times 15]}$, and the last ${[3 \times 30]}$. 
As CNNs do not include dynamic information, we smooth the CNN position estimates using a linear Kalman filter (LKF) with a constant velocity dynamic model to assure a fair comparison. 
A subsequent PF is not applicable in this case, as the trained CNN only produces a position estimate from the recorded data directly.

\section{Feature Analysis}
\label{sec:featan}

As mentioned in Sec.~\ref{sec:infex}, the effects of the different extracted features on positioning accuracy are unclear in many cases. 
Also, unlike for the TOFs and similar measures \cite{WinTheo}, to our knowledge no theoretical positioning error bounds for them have been deducted.

\subsection{Spatial Analysis}

\begin{figure*}[tb!]
\centering
    \begin{subfigure}[b]{0.27\linewidth}
        \centering
        \includegraphics[height=3.8cm, trim=0 53 90 0, clip]{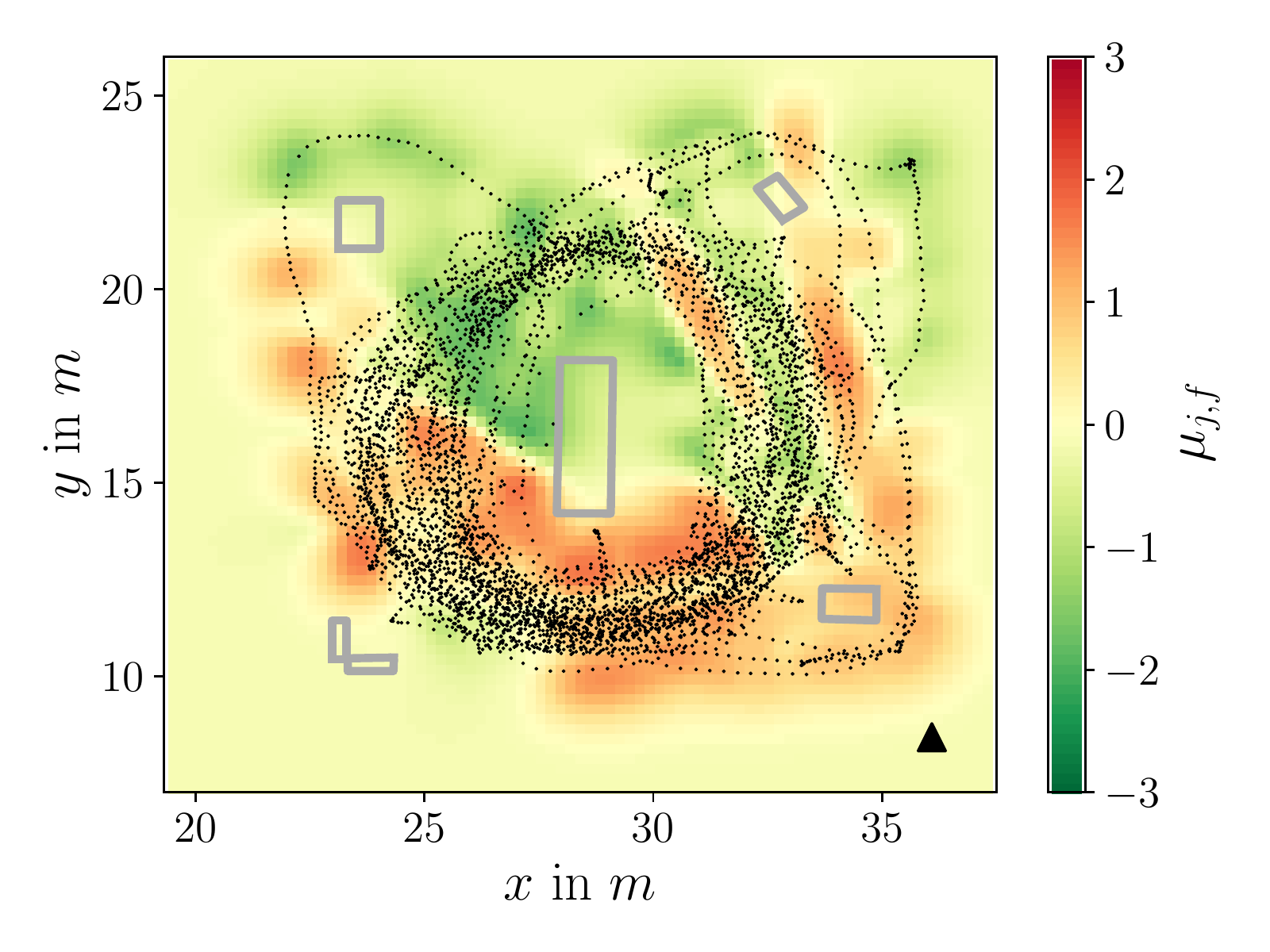}
    \end{subfigure}
    \hspace{-0.25cm}
    \begin{subfigure}[b]{0.22\linewidth}
        \centering
        \includegraphics[height=3.8cm, trim=55 53 90 0, clip]{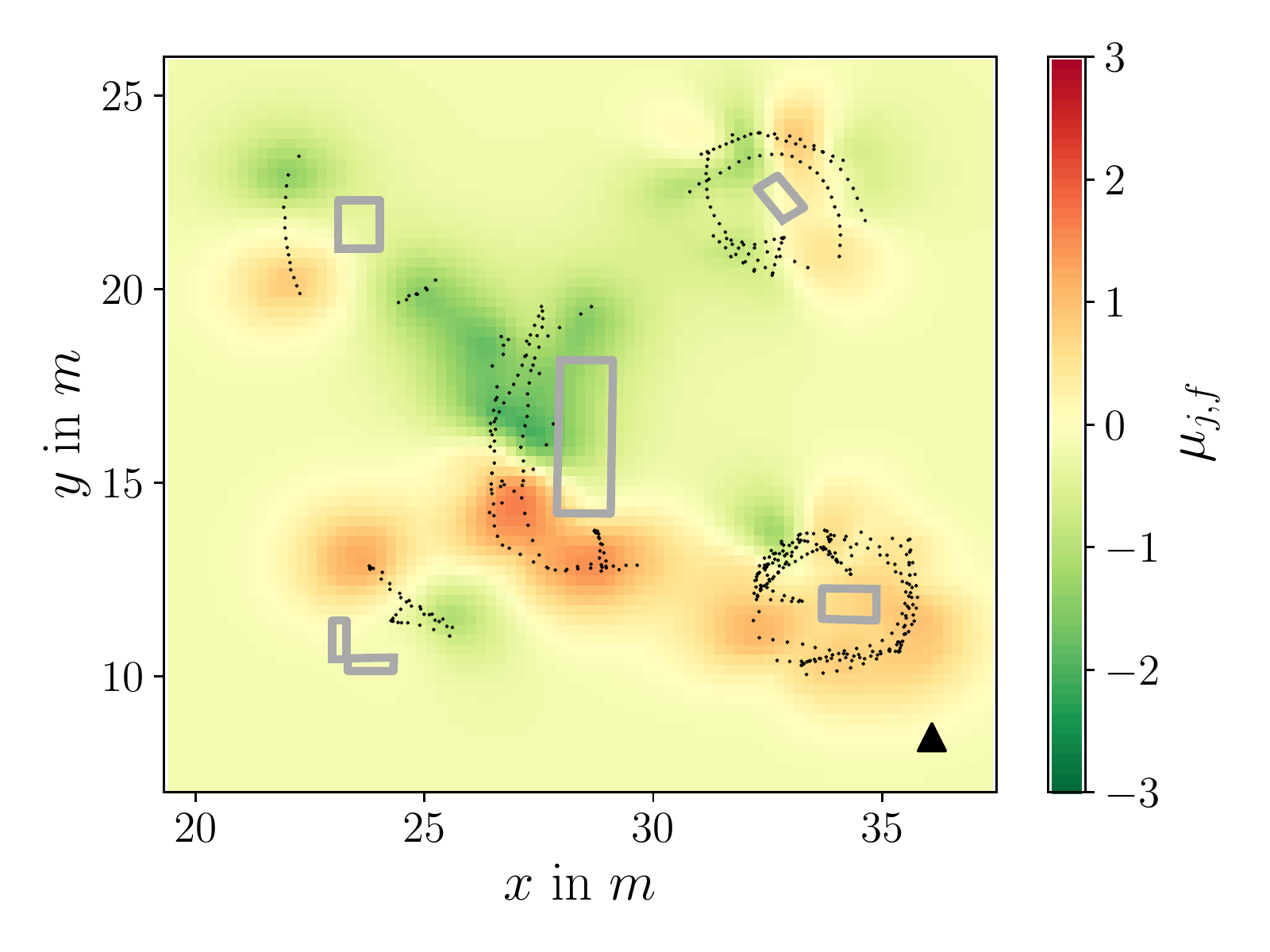}
    \end{subfigure}
    \begin{subfigure}[b]{0.22\linewidth}  
        \centering 
        \includegraphics[height=3.8cm, trim=55 53 90 0, clip]{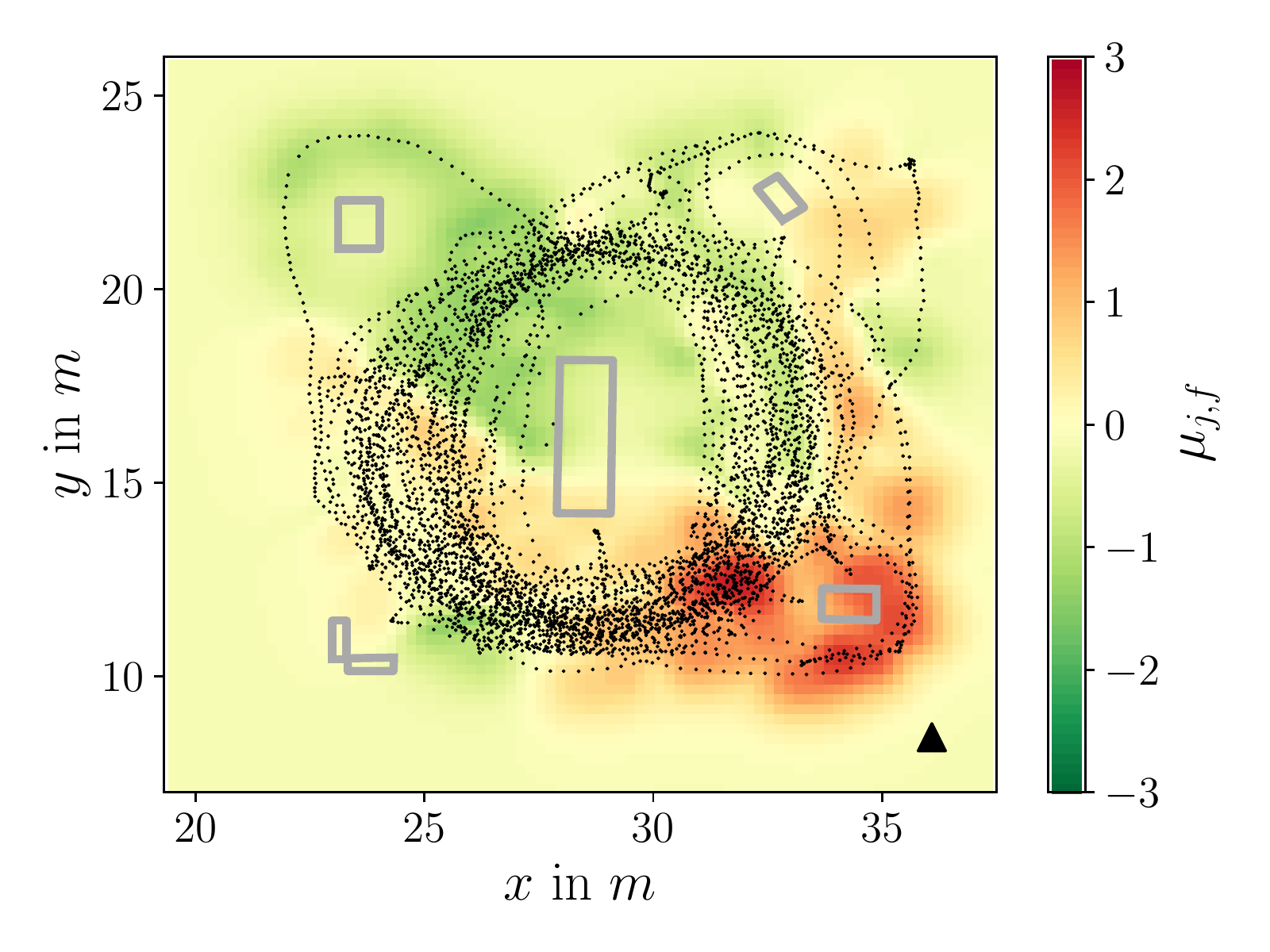}
    \end{subfigure}
    \begin{subfigure}[b]{0.27\linewidth}  
        \centering 
        \includegraphics[height=3.8cm, trim=55 53 0 0, clip]{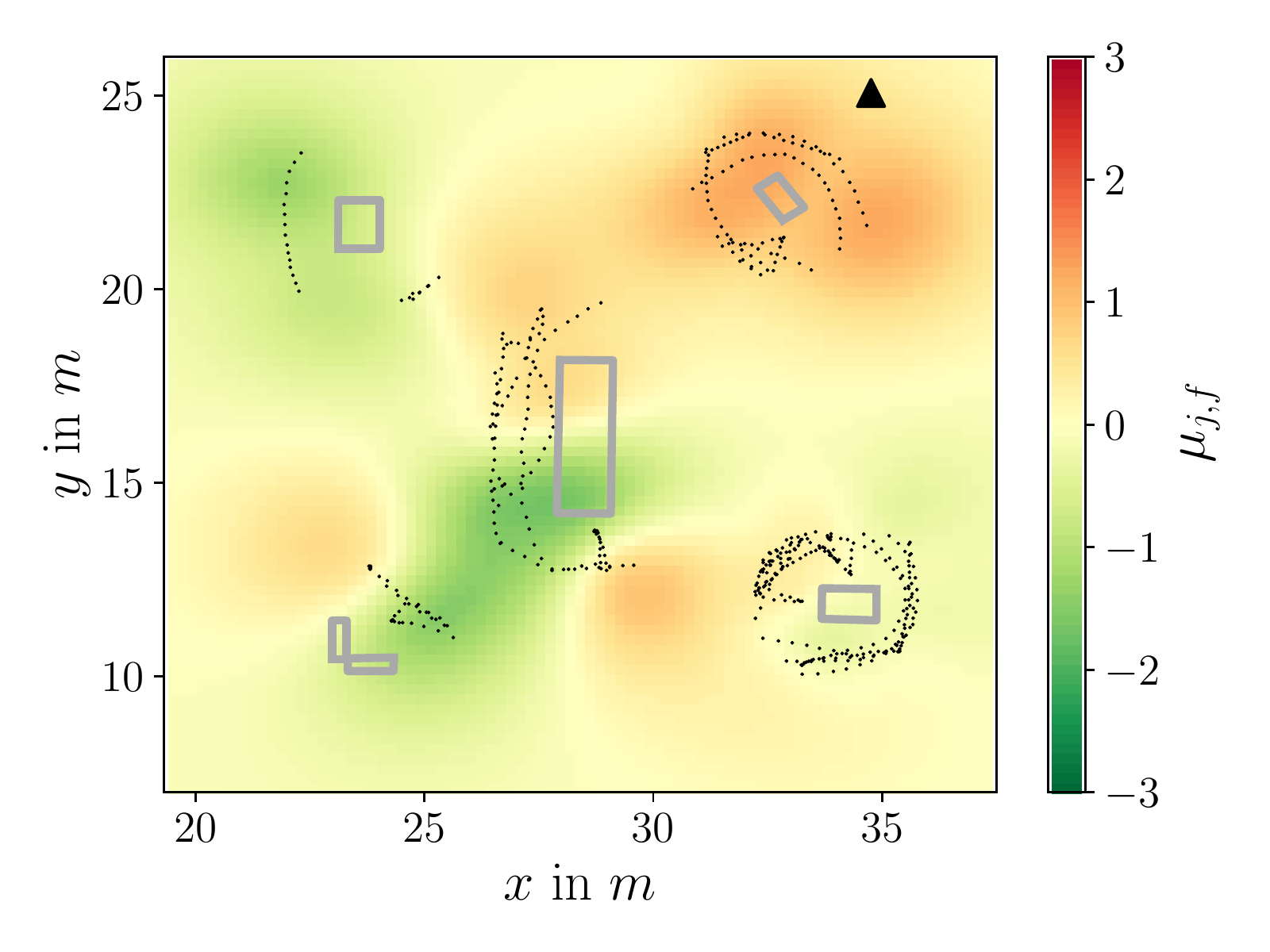}
    \end{subfigure}

    \begin{subfigure}[b]{0.27\linewidth}
        \centering
        \includegraphics[height=4.45cm, trim=0 0 90 0, clip]{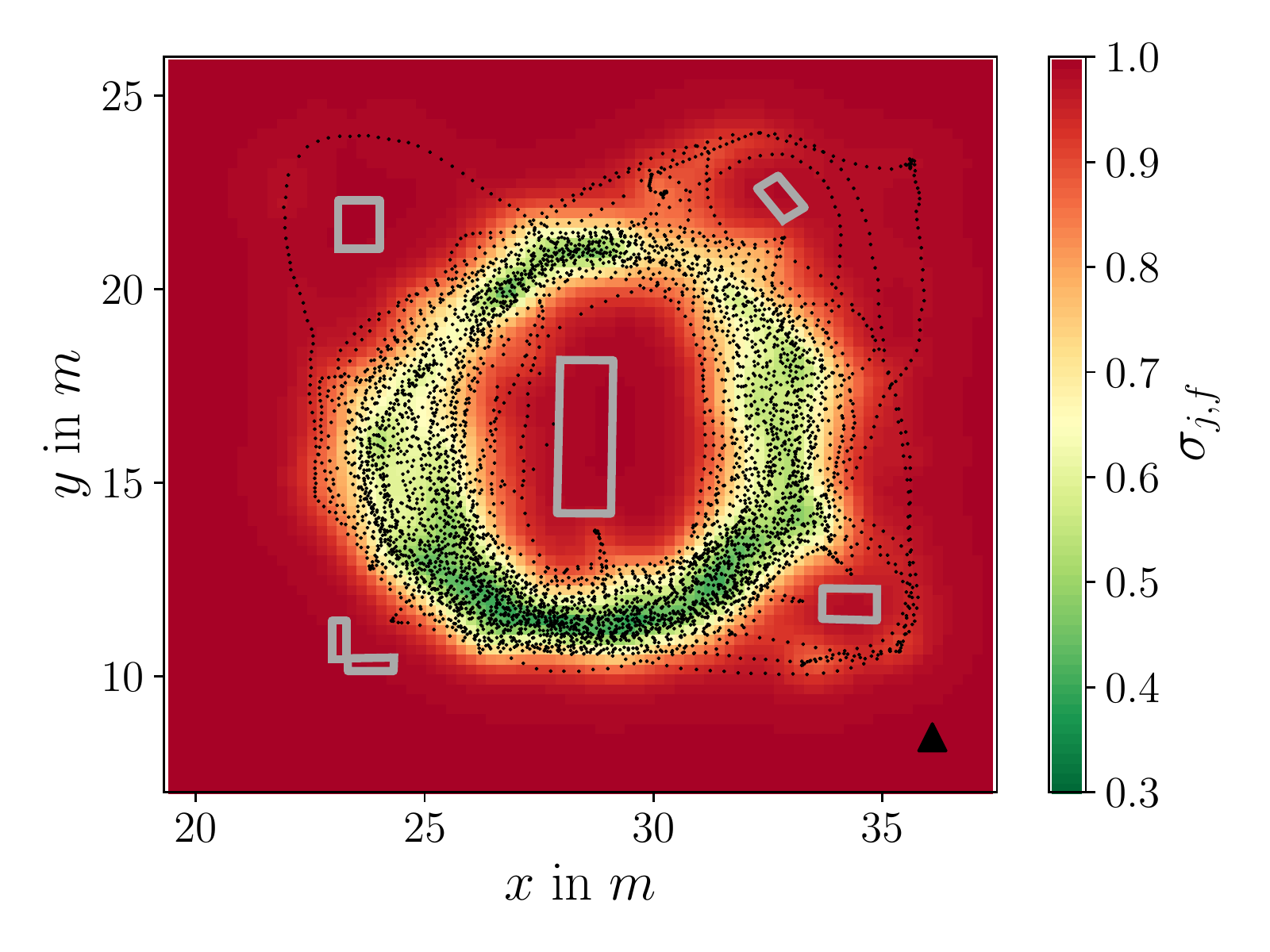}
        \caption{KUR, full dataset}  
        \label{fig:spatkurt1} 
    \end{subfigure}
    \hspace{-0.25cm}
    \begin{subfigure}[b]{0.22\linewidth}
        \centering
        \includegraphics[height=4.45cm, trim=55 0 90 0, clip]{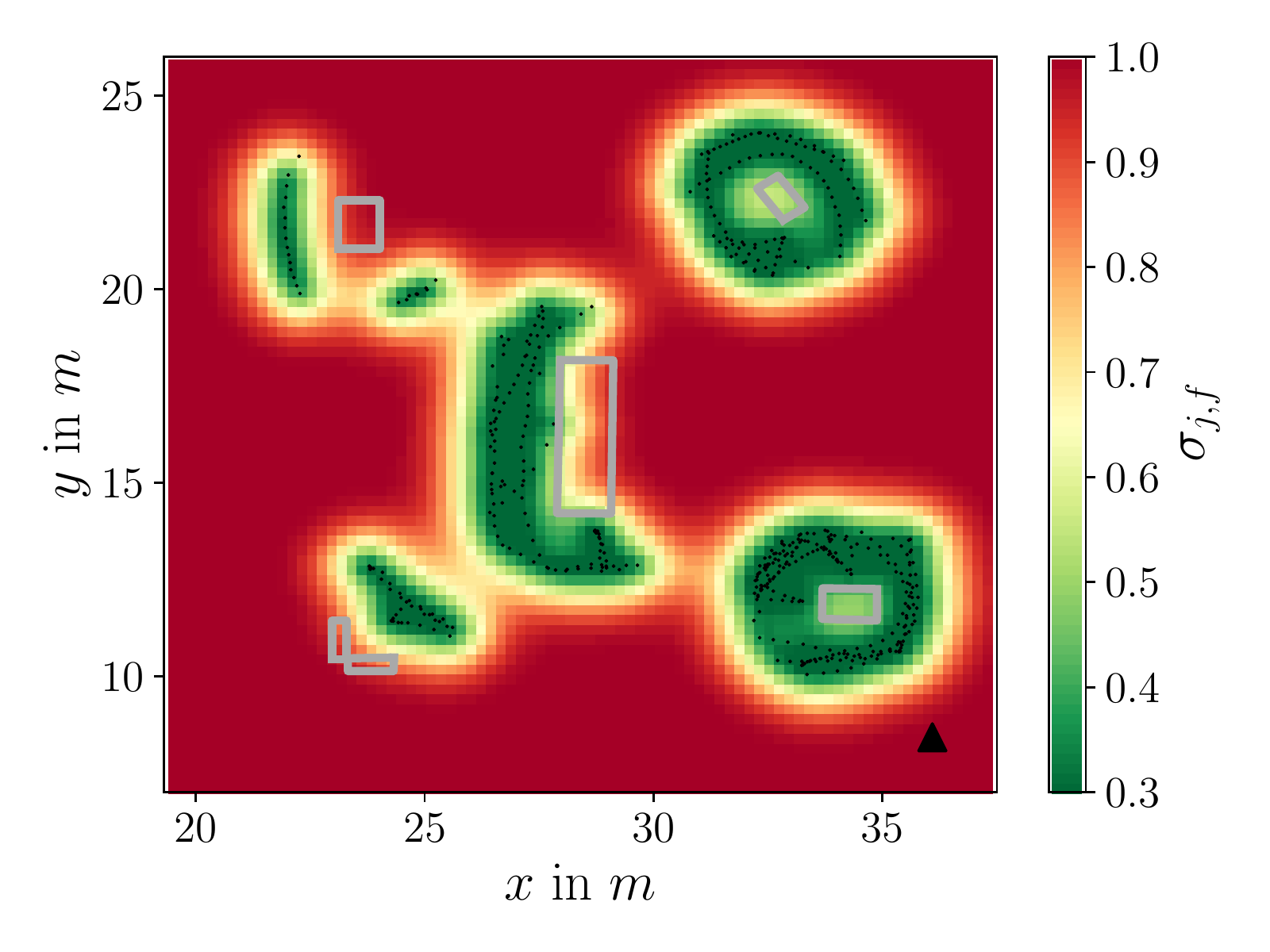}
        \caption{KUR, sparse dataset.}  
        \label{fig:spatkurt2} 
    \end{subfigure}
    \begin{subfigure}[b]{0.22\linewidth}  
        \centering 
        \includegraphics[height=4.45cm, trim=55 0 90 0, clip]{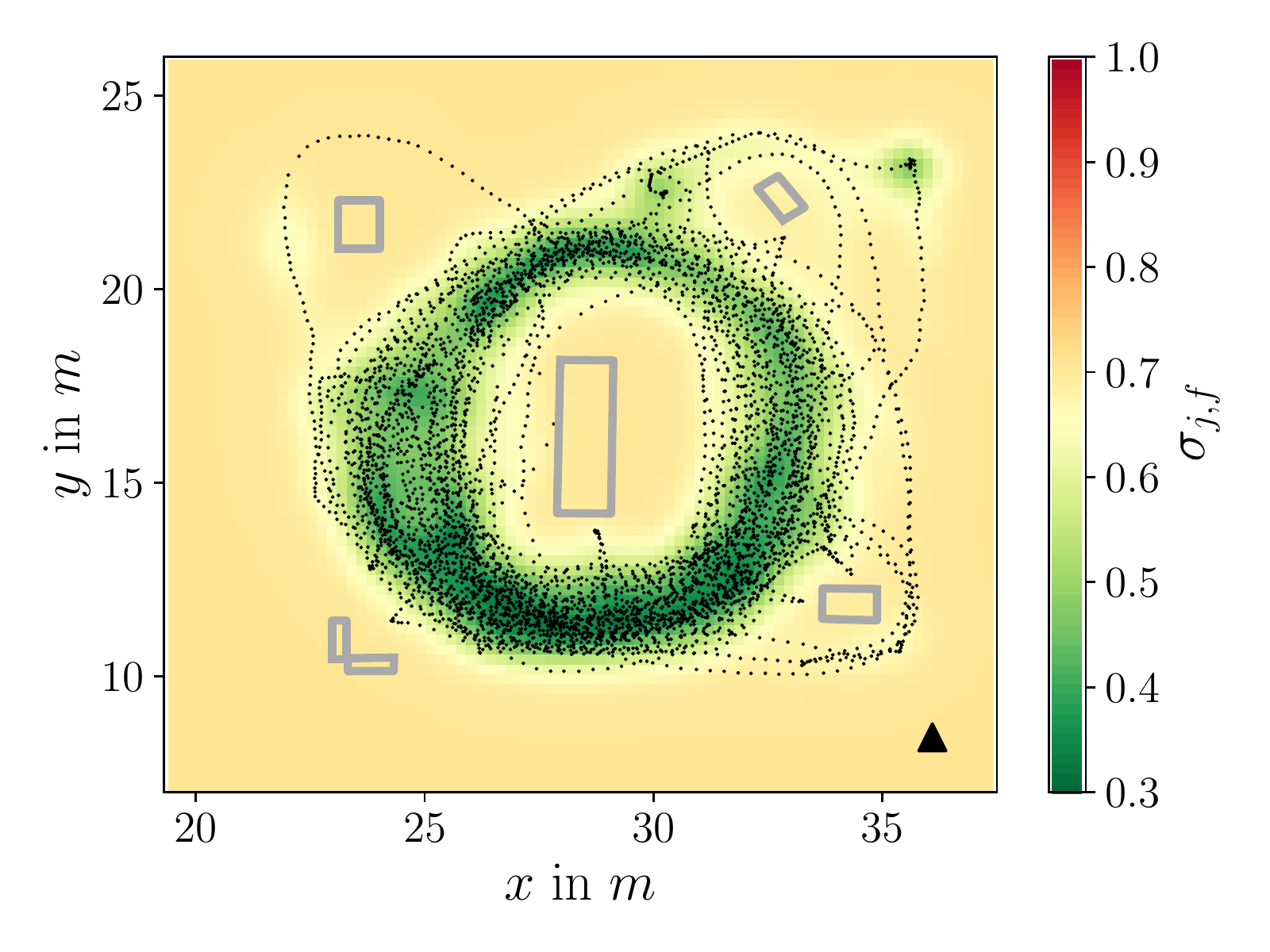}
        \caption{AE $1$, full dataset.} 
        \label{fig:spatlat1}
    \end{subfigure}
    \begin{subfigure}[b]{0.27\linewidth}  
        \centering 
        \includegraphics[height=4.45cm, trim=55 0 0 0, clip]{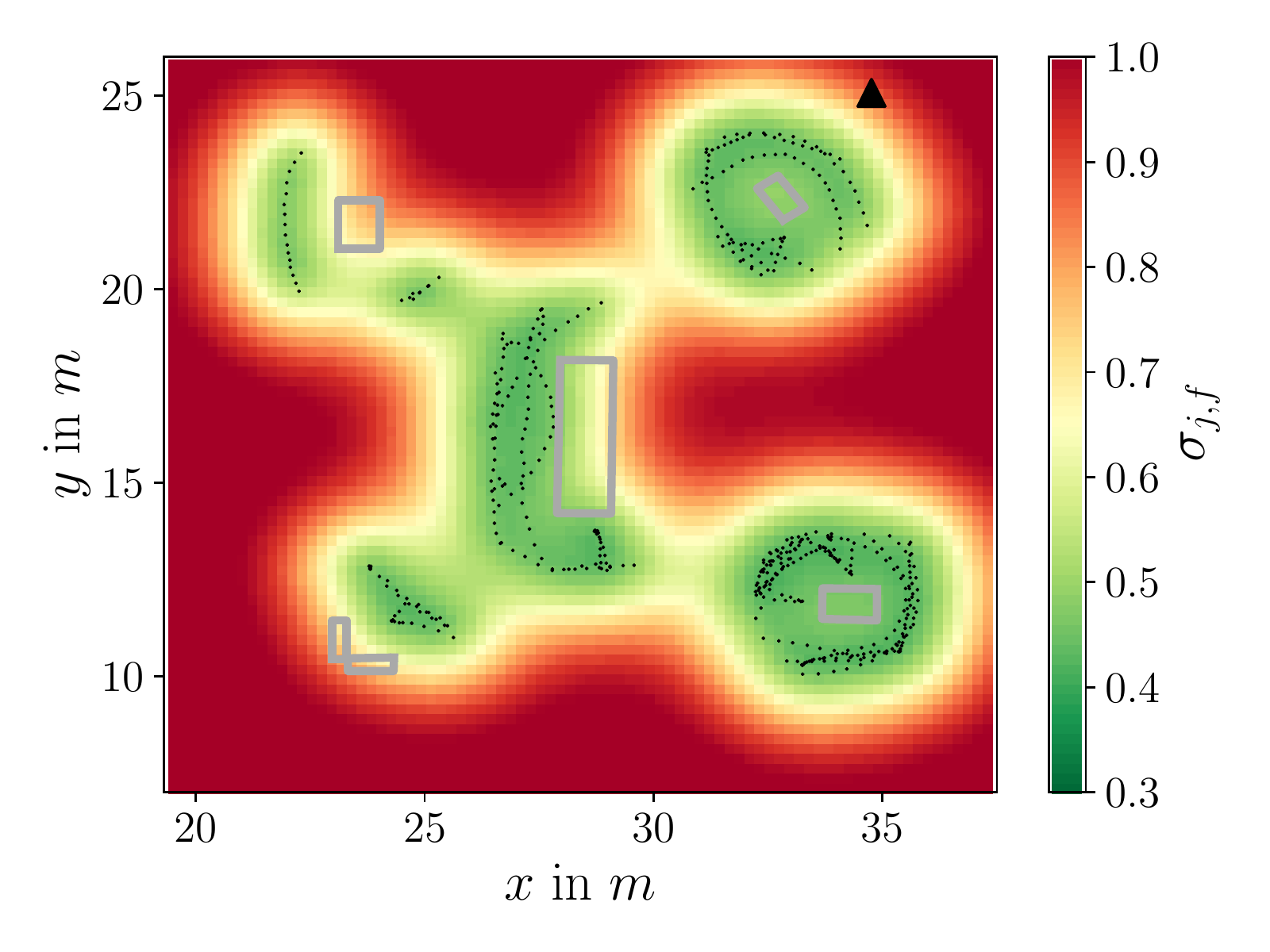}
        \caption{AE $2$, sparse dataset.} 
        \label{fig:spatlat2}
    \end{subfigure}    
    
    \caption{Means (top) and standard deviations (bottom) of Gaussian processes modeling the spatial distribution of different features based on both the \textit{full} and \textit{sparse} datasets.
  The anchor position is denoted by the black triangle, the positions of the FPs $\mathcal{Z}^{j,f}$ are visualized as black dots. Object positions are indicated with grey boxes.}
    \label{fig:traj_errors}
\end{figure*}

To interpret the spatial behavior of the GPs, we conduct a spatial analysis of a representative selection. 
Note that all GPRs work on scaled data, so that the visualization is in arbitrary units.

First, we consider the \textbf{magnitude kurtosis (KUR)} (see Figs.~\ref{fig:spatkurt1} and ~\ref{fig:spatkurt2}). 
For the full dataset (Figs.~\ref{fig:spatkurt1}), GPR learns a detailed spatial distribution for $\mu_{j,f}$ with distinct areas of high (red/orange color) and low (green color) values. 
Outside of the recording area, $\mu_{j,f}$ converges to a global mean of $0$ due to the scaling. 
Based on this distribution different areas can be distinguished from each other, e.g., the area behind the absorber wall around $[x, y] = [27, 16]$ from the area between the absorber wall and metal shelves around $[30, 12]$. 
$\sigma_{j,f}$ (Fig.~\ref{fig:spatkurt1}, top) increases rapidly outside of the areas with dense recordings. 
Considering the sparse dataset with the same feature (Figs.~\ref{fig:spatkurt2}),  $\mu_{j,f}$ only diverges from $0$ in the areas with data recordings, correspondingly $\sigma_{j,f}$ increases rapidly outside of these areas. 
Thus, in these areas mostly uniform weights are assigned by the GP (see Sec.~\ref{subsec:obsmod}). Hence, the tracking filter may rely on the TOF as a primary information source.

Second, we consider the \textbf{autoencoder features}. 
We visualize GPs of $2$ from the $8$ latent variables generated with CNN-AE-P for different anchors, one for a full and one for a sparse dataset. 
Due to randomized initialization of the AE, the ordering within the latent space is arbitrary. 
As with KUR, for the full dataset (Fig.~\ref{fig:spatlat1}), a detailed spatial distribution of $\mu_{j,f}$ is generated within the recording area.
However, $\sigma_{j,f}$ increases less rapidly, indicating a longer scale parameter of the kernel. 
If we compare this spatial distribution with Fig. ~\ref{fig:spatkurt1}, it becomes obvious that the latent variable encodes similar information to KUR, albeit with less diversity in the top right corner and significantly higher values of $\mu_{f,j}$ at the metal shelf around $[35, 12]$. 
Hence, there is obvious redundancy and therefore exploiting both features may be detrimental to the tracking performance.

Fig.~\ref{fig:spatlat2} visualizes the GP trained on the sparse dataset for another feature recorded with another anchor. 
In comparison with Figs.~\ref{fig:spatkurt1} and ~\ref{fig:spatlat1}, this feature allows for distinction between other areas, e.g., splitting the area below the absorber wall around $[30, 12]$ into subareas with high and low values of $\mu_{j,f}$. 
While the area around the shelves at $[35, 12]$ does not exhibit a significant variation value of $\mu_{j,f}$, this is the case for KUR and the other latent variable (Figs. \ref{fig:spatkurt1} and \ref{fig:spatlat1}). 
For the sparse dataset (Fig.~\ref{fig:spatlat2}), the different scaling of the kernels again results in $\sigma_{f,j}$ to increase less rapidly than for KUR. $\mu_{f,j}$ converges to the global mean (yellow color) away from observations, however this increase is less rapid due to the scale parameter of the kernel. 
In comparison to Figs.~\ref{fig:spatkurt2} and~\ref{fig:spatlat1}, there is a high difference in the means below the absorber walls around $[27, 14]$, so that the feature provides unique spatial information in this area. 

These exemplary results imply that a combination of both (different) features and anchors is beneficial, as they may embed unique characteristic spatial information. However, it is not obvious which possible subset of the features to employ for tracking to balance between redundancy and variation. 
Unlike ToA and deterministic MPCs~\cite{LeitingerPINFMP}, both the propagation and AE features do not have a straightforward relation to the position information they imply. Hence, a straightforward derivation of their contribution to the positioning accuracy is impossible. Therefore, we conduct a gridsearch over all possible subsets of the propagation features and AE features.

\subsection{Feature Selection}
\label{sec:exp:feature_selection}

To select the most characteristic feature combinations that embed the best possible spatial information, we evaluated the positioning accuracy with our test dataset to obtain the best subsets. 
We found that both feature types are beneficial to the positioning performance, with the AE features outperforming the propagation features. 
While the best performing propagation features are RKF and KUR, CNN-AE-P with 8 latent variables yields the best positioning accuracy. Using a subset of 3-6 variables yielded the best results overall, hence avoiding redundancy.
To examine whether the positioning accuracy can be further improved when we combine both feature types, we conducted a feasibility test by taking the combination of the best AE features and all possible subsets of propagation features. 
No combination, however, did improve the positioning accuracy further.

\section{Positioning Results}
\label{sec:posres}

After deriving and selecting characteristic features that encode important spatial information, we now evaluate if they can be employed to enhance positioning results with our proposed framework in comparison with the baseline methods. 
For quantitative evaluation of the position accuracy we consider the statistics of the absolute position error (APE). Table~\ref{tab:AEs} lists all the positioning results.

\begin{figure}[b!]
    \centering
    \includegraphics[width=0.8\linewidth]{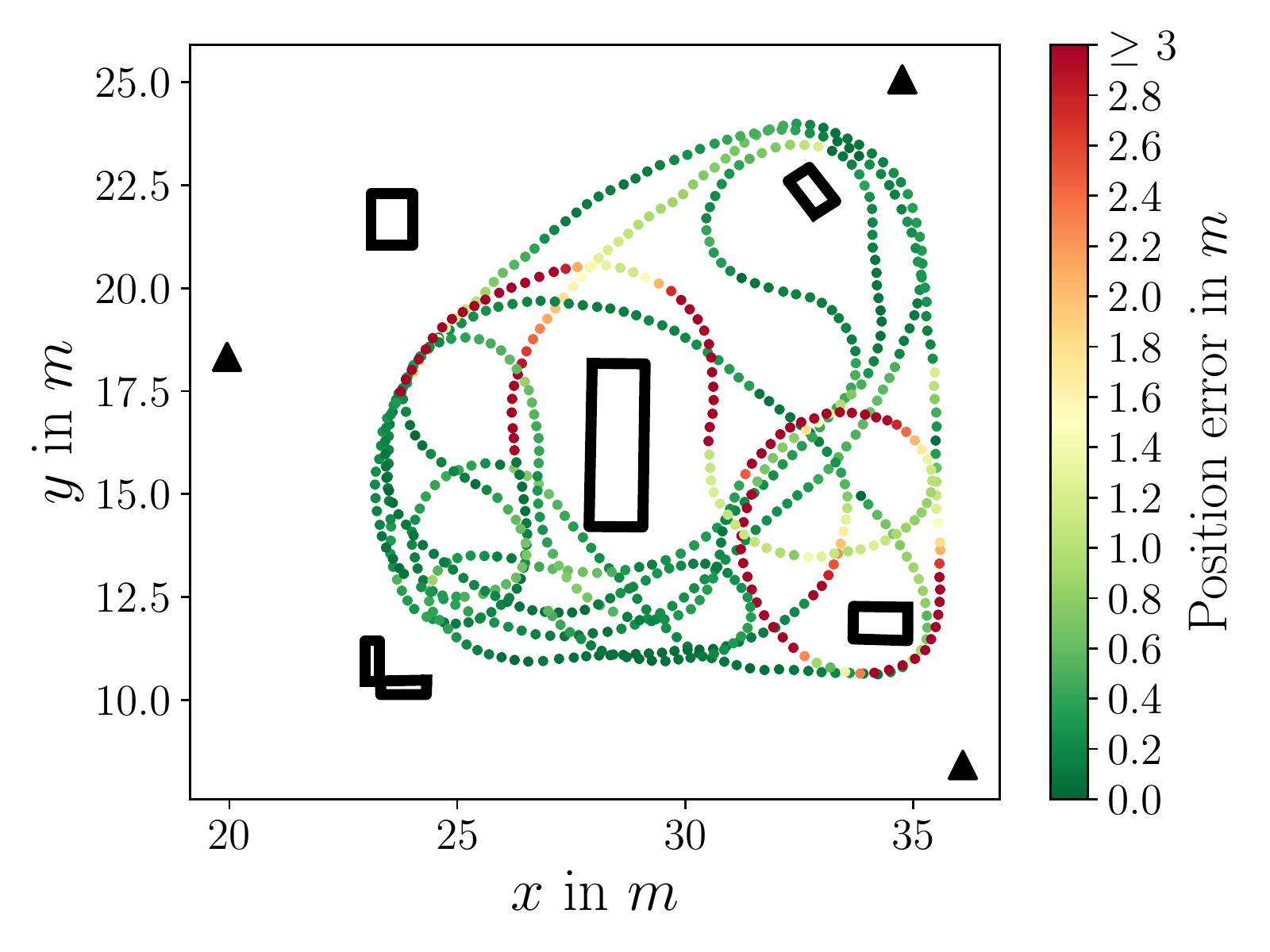}
    \caption{Position error over the trajectory for the EMI approach.}
    \label{fig:emi_traj3_anchs}
\end{figure}
\begin{figure*}[tb!]
    \centering
    \begin{subfigure}[b]{0.3\linewidth}
        \centering
        \includegraphics[height=4.3cm, trim=0 50 90 0, clip]{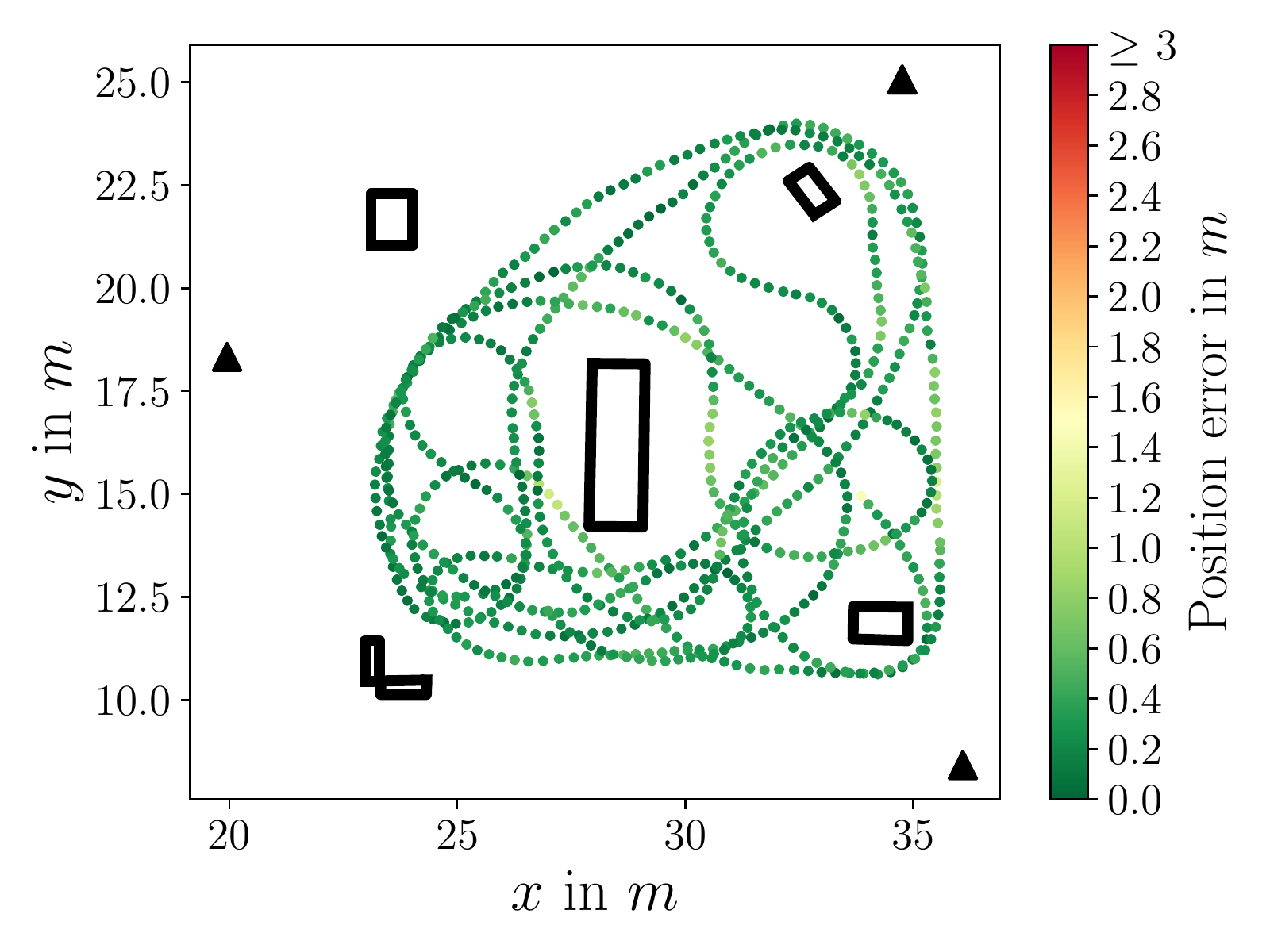}
        \caption{CNN-FP, full dataset.}  
        \label{fig:traj_cnn_full}
    \end{subfigure}
    \begin{subfigure}[b]{0.3\linewidth}  
        \centering 
        \includegraphics[height=4.3cm, trim=65 50 90 0, clip]{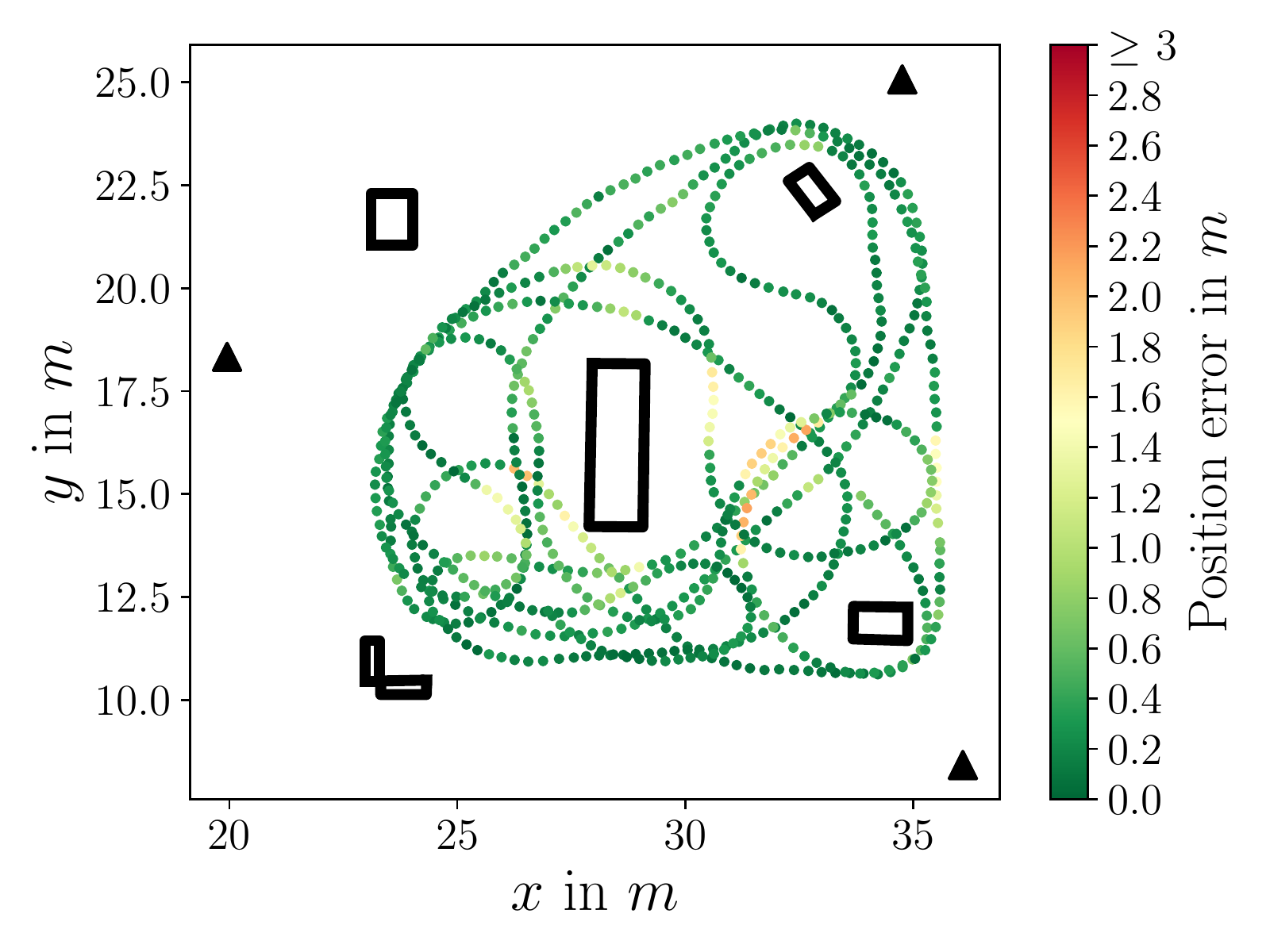}
        \caption{GPR-P, full dataset.} 
        \label{fig:traj_gpr_full_cf}
    \end{subfigure}
    \hspace{-0.5cm}
    \begin{subfigure}[b]{0.36\linewidth}
        \centering
        \includegraphics[height=4.3cm, trim=65 50 0 0, clip]{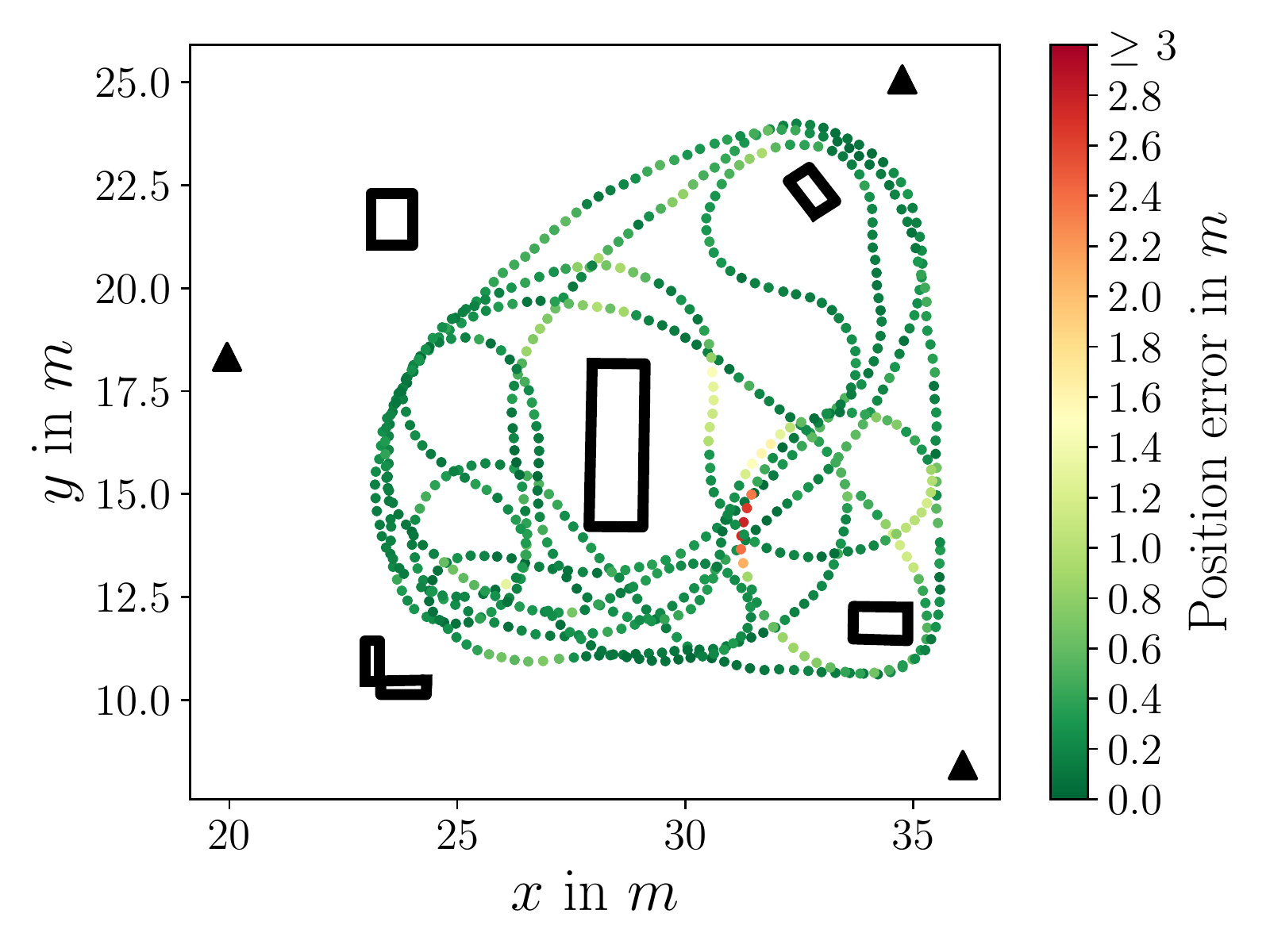}
        \caption{GPR-AE, full dataset.}  
        \label{fig:traj_gpr_full_ae}
    \end{subfigure}
    
    \begin{subfigure}[b]{0.3\linewidth}
        \centering
        \includegraphics[height=5.05cm, trim=0 0 90 0, clip]{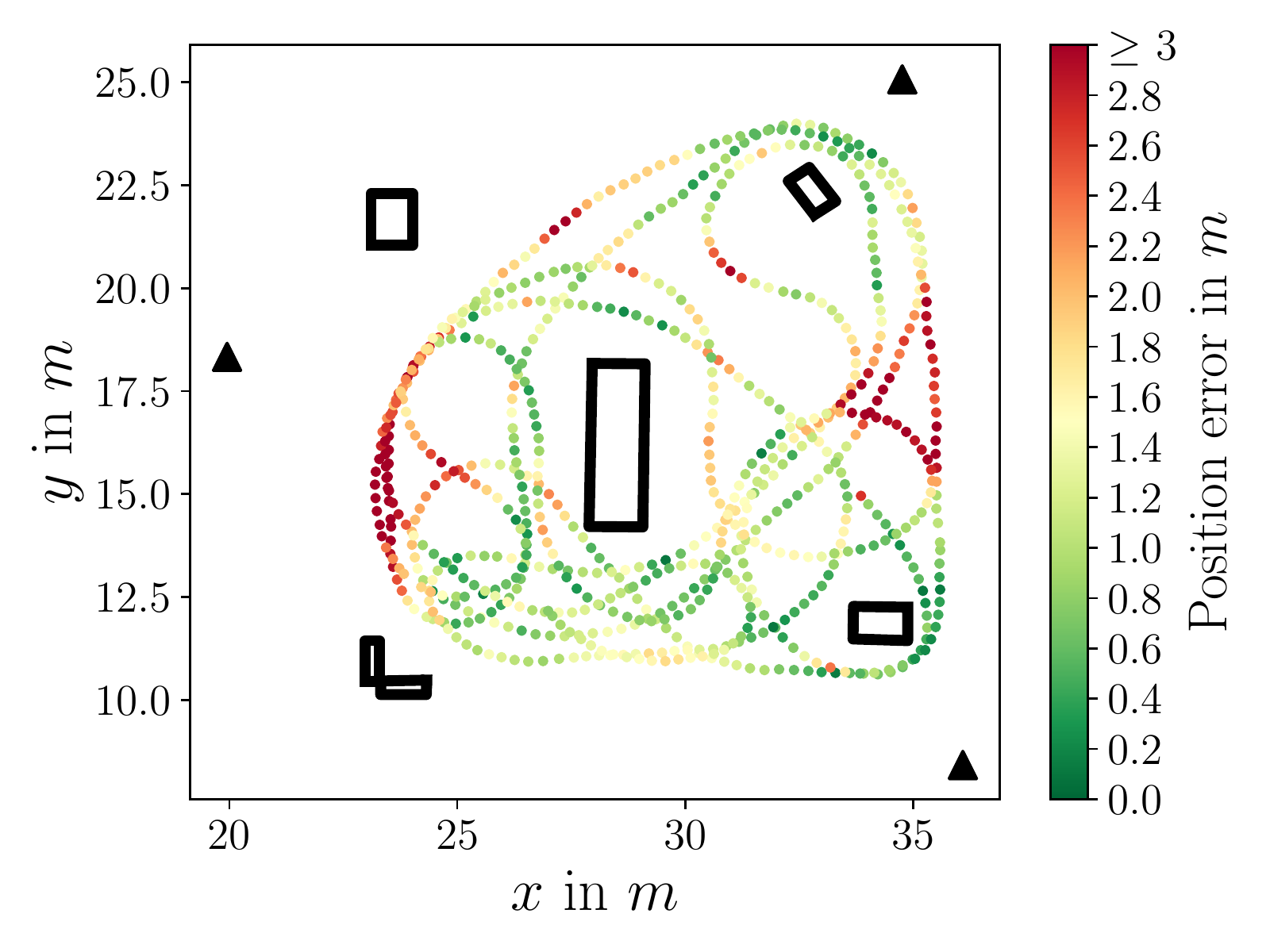}
        \caption{CNN-FP, sparse dataset.}  
        \label{fig:traj_cnn_sparse}
    \end{subfigure}
    \begin{subfigure}[b]{0.3\linewidth}  
        \centering 
        \includegraphics[height=5.05cm, trim=65 0 90 0, clip]{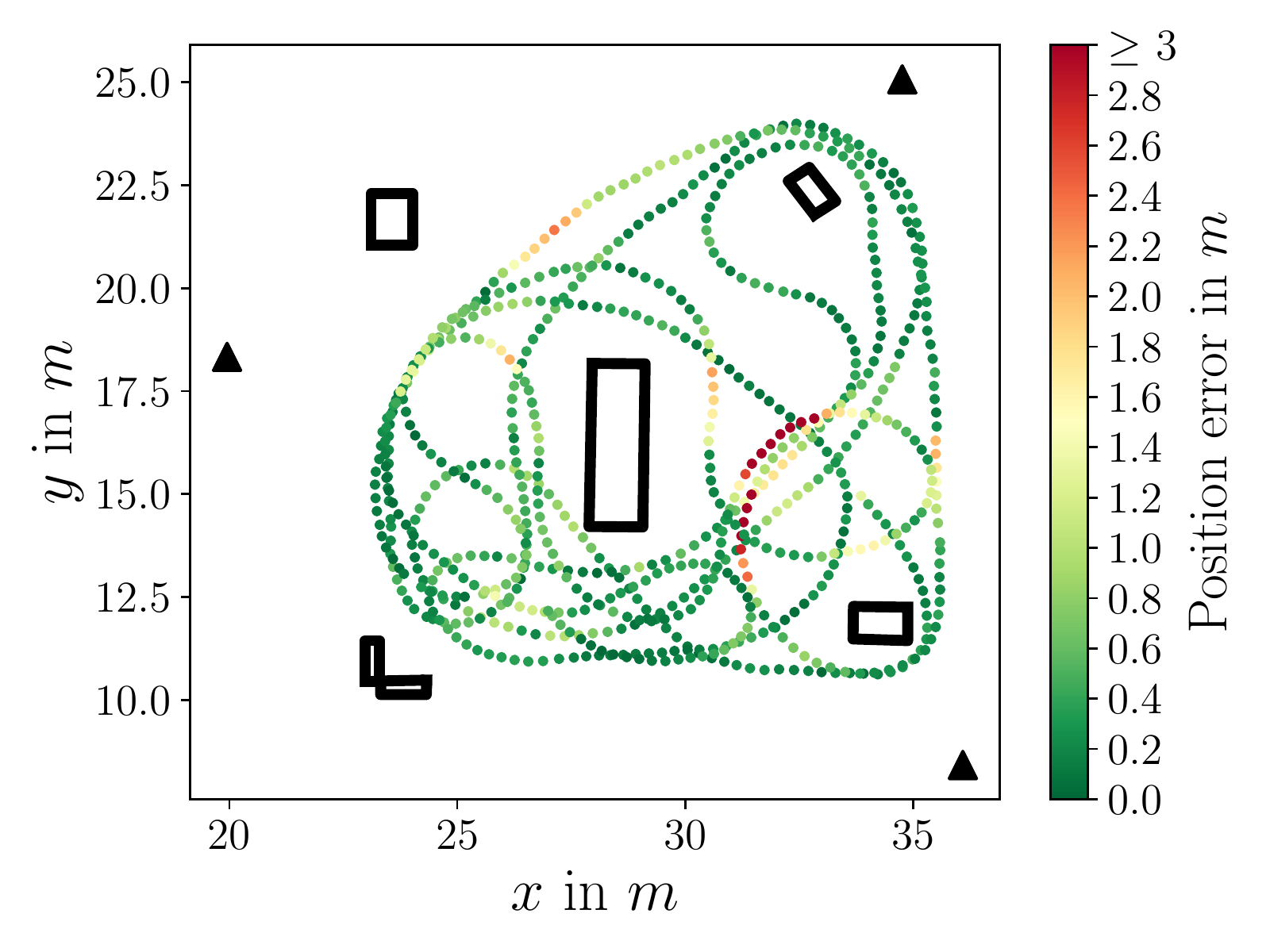}
        \caption{GPR-P, sparse dataset.} 
        \label{fig:traj_gpr_sparse_cf}
    \end{subfigure}
    \hspace{-0.5cm}
    \begin{subfigure}[b]{0.36\linewidth}
        \centering
        \includegraphics[height=5.05cm, trim=65 0 0 0, clip]{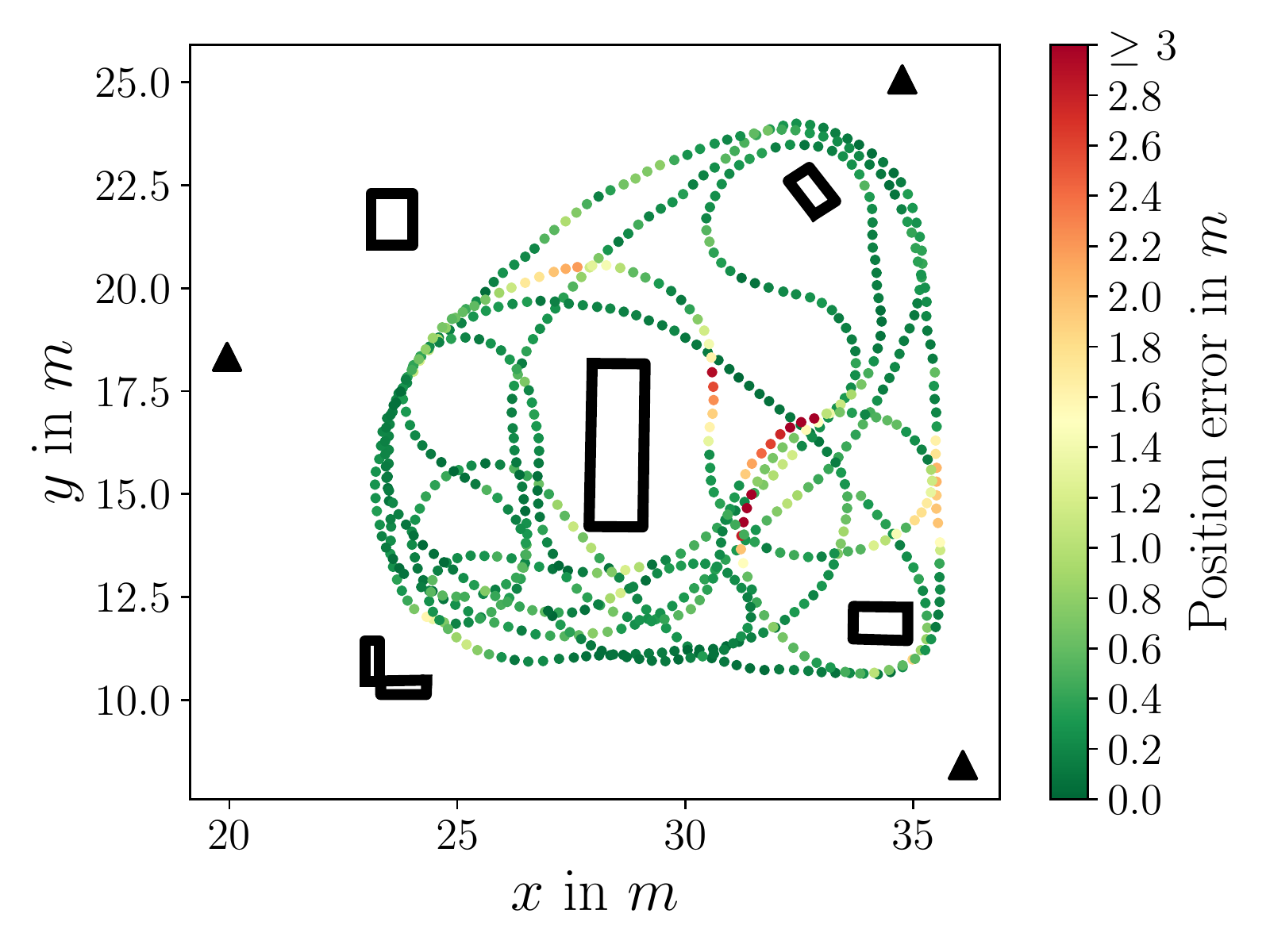}
        \caption{GPR-AE, sparse dataset.} 
        \label{fig:traj_gpr_sparse_ae}
    \end{subfigure}
    
    \caption{Positioning errors of the FP approaches on the full (top row) and sparse (bottom row) datasets. 
    The anchor positions are indicated with black triangles, while the black boxes indicate the positions of objects within the environment.}
    \label{fig:traj_errors}
\end{figure*}

\begin{figure}[tb]
    \centering
    \includegraphics[width=0.9\linewidth]{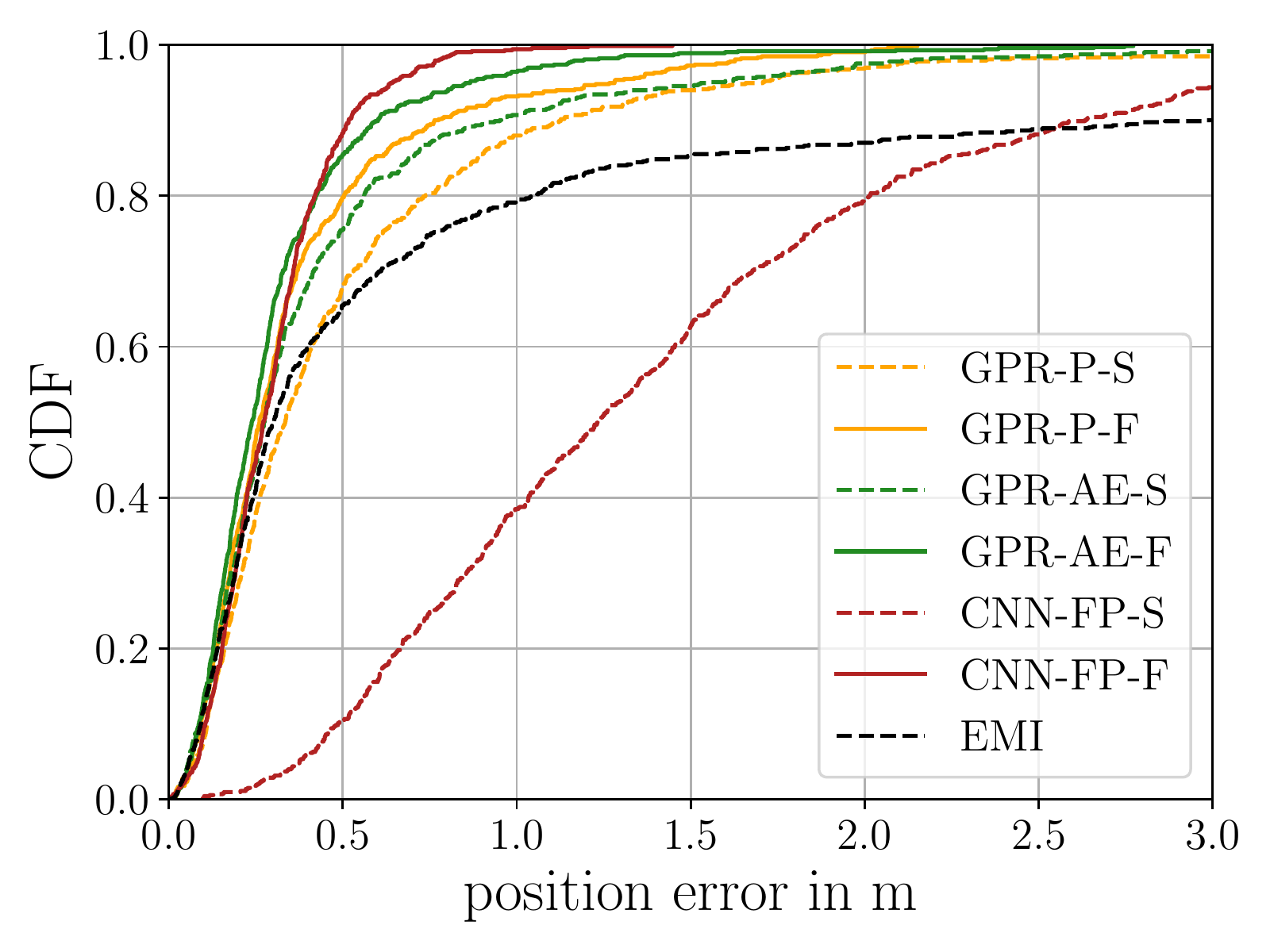}
    \caption{Position error CDFs for 3 Anchors.}
    \label{fig:error_cdf_3}
\end{figure}

Fig.~\ref{fig:emi_traj3_anchs} shows the results of the EMI approach in the proposed scenario: 
Stable tracking with APEs in the low decimeter range (green color) is possible in most parts of the environment. 
However some areas are problematic, leading to a complete failure of the tracking filter, resulting in APEs in the higher meter range (yellow to red color). 
Interestingly, outside of these zones, the tracking filter converges again to a low APE. 
Therefore, due to low LOS anchor availability in problematic areas additional spatial information is required to maintain accurate tracking. 
Although a relatively low error median (MED) of $31.2$\si{cm} can be achieved (see Fig.~\ref{fig:error_cdf_3}, Table~\ref{tab:AEs}), significant outliers occur in the areas with low LOS coverage that result in a high mean absolute error (MAE) of $1.00$\si{m} and C95 of  $4.79$\si{m}. 
In relevant applications, such as robot or automatic industrial vehicle navigation or tool tracking, such deviations may lead to serious problems such as crashes with structures or the loss of equipment. 
However, these errors can be avoided by FP methods (i.e., the baseline CNN-FP and our proposed GPR method) as they exploit additional spatial information contained in the FP database.

\begin{table}[t!]
    \centering
    \caption{Mean (MAE), Median (MED), 75th and 95th percentile (C75 / C95) of the  APE  in \si{m}.}
    \begin{tabular}{c||c| c| c| c |c}
    & MAE & MED & C75 & C95 & ref. \\
    \hline
         EMI & 1.00 & 0.312 & 0.778 & 4.79 & Fig.~\ref{fig:emi_traj3_anchs} \\
         CNN-FP-F & 0.300 & 0.271 & 0.385 & 0.639 & Fig.~\ref{fig:traj_cnn_full}\\
         CNN-FP-S & 1.39 & 1.23 & 1.85 & 3.08 & Fig.~\ref{fig:traj_cnn_sparse}\\
         GPR-P-F & 0.409 & 0.276 & 0.467 & 1.35 & Fig.~\ref{fig:traj_gpr_full_cf}\\
         GPR-P-S & 0.634 & 0.366 & 0.727 & 2.22 & Fig.~\ref{fig:traj_gpr_sparse_cf}\\
         GPR-AE-F & 0.333 & 0.245 & 0.390 & 0.906 & Fig.~\ref{fig:traj_gpr_full_ae}\\
         GPR-AE-S & 0.502 & 0.289 & 0.535 & 1.755 & Fig.~\ref{fig:traj_gpr_sparse_ae}\\ 
    \end{tabular}

    \label{tab:AEs}
\end{table}

For the full dataset CNN-FP can maintain accurate tracking in  the complete environment (Fig.~\ref{fig:traj_cnn_full}): 
Due to full data availability, the CNN estimates all positions accurately, while potential outliers are smoothed out by the dynamics model of the LKF. 
This is also seen in the error statistics (Fig.~\ref{fig:error_cdf_3}, Tab.~\ref{tab:AEs}); the MAE is $0.300$\si{m}, errors below $38.5$\si{cm} can be achieved in $75~\%$ of the cases. 
However, a massive deterioration of the positioning accuracy occurs for the \textit{sparse} dataset: CNN-FP lacks the necessary data in large parts of the environment and cannot properly extrapolate (Fig.~\ref{fig:traj_cnn_sparse}). 
As CNN-FP does not rely on observation modeling and only produces non-stochastic position estimates, the LKF tracking algorithm cannot adapt to this change in reliability. 

Hence, in a typical application setting this reliability may remain undetected and may lead to system failure. This is also reflected in the error statistics (Fig.~\ref{fig:error_cdf_3}), where both the error median and $95$th percentile (C95) increase significantly to $1.23$\si{m} abd $3.08$\si{m}. Also, the MAE and C95 deteriorate to $1.39$\si{m} and $3.08$\si{m}. 
Note that in comparison with Fig.~\ref{fig:emi_traj3_anchs}, CNN-FP suffers from these outliers in areas where enough LOS information is present to enable stable tracking for EMI. Our method compensates the shortcomings of both baselines by employing a statistical observation modeling that combines information sources, i.e., the ToF and FP dataset, in a PF.

Based on our feature selection, we compare the best performing propagation feature GPs (GPR-P) and AE feature GPs (GPR-AE).
Both GPR-P-F and GPR-AE-F significantly enhance the position accuracy over the EMI approach for the full dataset (Figs.\ref{fig:traj_gpr_full_cf}, \ref{fig:traj_gpr_full_ae}).
In general, the GPR-AE-F performs better than GPR-P-F and yields a MAE of $0.330$\si{m} and MED of $0.246$\si{m}, which is lower than CNN-FP-F. 
However, GPR-AE-F returns outliers with higher errors (C95 of $0.906$\si{m}) than CNN-FP-F. 
This may be due to a more problem-specific and detailed feature selection in the CNN, while our method compresses the information by feature extraction.

In contrast to the results on the full dataset, GPR-AE-F, unlike CNN-FP, show it full potential on the sparse dataset (Fig.~\ref{fig:traj_gpr_sparse_cf}, \ref{fig:traj_gpr_sparse_ae}): 
In areas where no data recordings are available, the observation model components of the GP models produce high variance estimates, so that the estimated log-likelihood weights are mostly uniform.
Thus, the majority of spatial information for the update of the PF is contributed by the LOS measurements, which, as seen in Fig.~\ref{fig:emi_traj3_anchs} are sufficient for accurate tracking.
Hence, the performance decrease of GPR-AE-S is much lower than for CNN-FP (Fig.~\ref{fig:error_cdf_3}), so that for GPR-AE-S, the MED only decreases to $0.287$\si{m}. 
However, in areas with low LOS anchor coverage significant outliers exist, most significantly in the area around $[32, 15]$. 
Here, neither the sparse FP dataset nor the obstructed TOFs can provide valid spatial information. 
This causes an increased C95 of $1.60$\si{m}. 
Thus, in general, with a spatially spare FP dataset, the position accuracy of GPR-AE-S is significantly higher than both baselines (CNN-FP-S and EMI). 

In essence, our approach maintains stable tracking even in large environments with mixed propagation conditions. We only require a low amount of data recorded specifically in areas that are dominated by (diffuse) MPC. This drastically reduces the effort of data collection and maintenance.

\section{Discussion and Limitations}

First, our \textit{feature extraction} resorts to suitable propagation features known from the literature and leaves out other types of features, e.g. time-frequency-domain features. While this of course is far from being complete we leave a more thorough discussion of usuable features to future work. 

Second, we assume \textit{statistical independence} between features in ~\eqref{eq:featslikel}. Such independence might not necessarily be guaranteed by how we extract features using the autoencoder. Adapted versions of the autoencoder, i.e., disentangled variational autoencoders, that assure independent features~\cite{NIPSVAE} may enforce this statistical independence. The statistical independence from the LOS component might be increased, e.g., by spectral subtraction.

Third, we might extract \textit{additional spatial information} as the features we extract are only applied to the magnitudes of the CM -- additional spatial information may be extracted from the phases. Also, we leave a study of how an additional pre-processing of the feature values might increase spatial consistency to future work. Moreover, the GPR in our framework builds upon stationary kernels, however, a domain-specific, non-stationary kernel might yield additional spatial information, like angular and radial dependence. 

Forth, for the \textit{real world application}, it might be easier to use multiple GPRs for the sparse dataset (e.g., one GP per object), to reduce computational complexity and database maintenance effort. Also, the influence of environment changes and consequent database adaption on positioning accuracy are of interest for operators of such positioning systems. How our approaches works on changed environmental conditions is yet to be analyzed more closely.

Fifth, we do not \textit{estimate weights} for the information sources, so that all of them are treated equally. An additional optimal weighting most probably increases the positioning accuracy and makes the use of more features more viable.
While we conducted a vast grid search to obtain a good representation, other methods such as evolutionary feature selection~\cite{Evolutionary} or reinforcement learning~\cite{ReinforcementFeature} may yield more optimal feature combinations and/or weights.

\section{Conclusion}
\label{sec:conclusion}

We presented a novel two-stage tracking framework for channel measurement (CM) based positioning. It builds upon the extraction of characteristic features (both propagation- and autoencoder-based (AE)) and applies Gaussian process regression (GPR) on recorded fingerprinting (FP) data by training individual Gaussian processes (GPs) for subsets of our features. The GPs model observation likelihoods that represent spatial information. Hence, our approach allows for an information fusion using, e.g., a particle filter (PF) with state-of-the-art dynamics modeling and resampling. Most importantly, our approach works with spatially sparse datasets.

We evaluate our methods on data of a realistic industrial environment. We investigate both a spatially dense and smaller and sparse training data sets. Unlike CNN-FP our method adapts well to the sparse dataset and yields significantly more accurate positions. We attribute this performance gain to the fact that the framework models the reliability of the observation likelihood based on stationary, distance-dependent kernels, and therefore can also effectively integrate the spatial information provided by EMI approaches and rely on it in areas without available data to maintain accurate positioning.

\section*{Acknowledgements}
This work was supported by the Bavarian Ministry for Economic Affairs, Infrastructure, Transport and Technology through the Center for Analytics-Data-Applications (ADA-Center) within the framework of “BAYERN DIGITAL II”. The authors also acknowledge the financial support by the Federal Ministry of Education and Research of Germany in the programme of “Souverän. Digital. Vernetzt.” Joint project 6G-RIC, project identification number: 16KISK020K.

\bibliographystyle{ieeetr}
\bibliography{references}
\newpage
\vfill

\end{document}